\documentclass[aps,pra,showkeys,onecite,twocolumn,citeautoscript,a4paper,superscriptaddress,10pt]{revtex4-1}

\usepackage[left=2.5cm, right=2.5cm, top=2.5cm, bottom=2.0cm]{geometry}

\usepackage[version=3]{mhchem}
\usepackage{graphicx} 
\usepackage[english]{babel}
\usepackage{siunitx}
\setcitestyle{super,bracket}

\begin{document}

\author{Simon Clark}
\affiliation{German Aerospace Center, Pfaffenwaldring 38–40, 70569 Stuttgart, Germany}
\affiliation{Helmholtz Institute Ulm, Helmholtzstra{\ss}e 11, 89081 Ulm, Germany}
\affiliation{SINTEF Industry, New Energy Solutions, Sem Saelands vei 12, 7034 Trondheim, Norway}
\author{Aroa R. Mainar}
\affiliation{CIDETEC, Basque Research and Technology Alliance (BRTA), P$^\circ$ Miram\'on, 196, Donostia-San Sebasti\'an 20014, Spain}
\author{Elena Iruin}
\affiliation{CIDETEC, Basque Research and Technology Alliance (BRTA), P$^\circ$ Miram\'on, 196, Donostia-San Sebasti\'an 20014, Spain}
\author{Luis C. Colmenares}
\affiliation{CIDETEC, Basque Research and Technology Alliance (BRTA), P$^\circ$ Miram\'on, 196, Donostia-San Sebasti\'an 20014, Spain}
\author{J. Alberto Bl\'azquez}
\affiliation{CIDETEC, Basque Research and Technology Alliance (BRTA), P$^\circ$ Miram\'on, 196, Donostia-San Sebasti\'an 20014, Spain}
\author{Julian R. Tolchard}
\affiliation{SINTEF Industry, New Energy Solutions, Sem Saelands vei 12, 7034 Trondheim, Norway}
\author{Zenonas Jusys}
\affiliation{Institute of Surface Chemistry and Catalysis, Ulm University, Albert-Einstein-Allee 47, 89081 Ulm, Germany}
\author{Birger Horstmann}\email{birger.horstmann@dlr.de}
\affiliation{German Aerospace Center, Pfaffenwaldring 38–40, 70569 Stuttgart, Germany}
\affiliation{Helmholtz Institute Ulm, Helmholtzstra{\ss}e 11, 89081 Ulm, Germany}
\affiliation{Faculty of Natural Sciences, Ulm University, 89081 Ulm, Germany}

\keywords{zinc-air batteries, aqueous near-neutral electrolyte, organic salt, thermodynamics, theory and validation}

\date{\today} 

\title{Designing Aqueous Organic Electrolytes for Zinc-Air Batteries: Method, Simulation, and Validation}

\begin{abstract}
Aqueous zinc-air batteries (ZABs) are a low-cost, safe, and sustainable technology for stationary energy storage. ZABs with pH-buffered near-neutral electrolytes have the potential for longer lifetime compared to traditional alkaline ZABs due to the slower absorption of carbonates at non-alkaline pH values. However, existing near-neutral electrolytes often contain halide salts, which are corrosive and threaten the precipitation of \ce{ZnO} as the dominant discharge product. This paper presents a method for designing halide-free aqueous ZAB electrolytes using thermodynamic descriptors to computationally screen components. The dynamic performance of a ZAB with one possible halide-free aqueous electrolyte based on organic salts is simulated using an advanced method of continuum modeling, and the results are validated by experiments. XRD, SEM, and EDS measurements of \ce{Zn} electrodes show that \ce{ZnO} is the dominant discharge product, and \textit{operando} pH measurements confirm the stability of the electrolyte pH during cell cycling. Long-term full cell cycling tests are performed, and RRDE measurements elucidate the mechanism of ORR and OER. Our analysis shows that aqueous electrolytes containing organic salts could be a promising field of research for zinc-based batteries, due to their \ce{Zn^{2+}} chelating and pH buffering properties. We discuss the remaining challenges including the electrochemical stability of the electrolyte components.
\end{abstract}

\maketitle

\section{Introduction}
The demand for high-performance energy storage is rapidly growing. Ambitious plans to increase the share of renewables in the electric grid and expand the market for electric vehicles (EVs) underscore the importance of developing next-generation electrochemical energy storage technologies. Currently this demand is overwhelmingly met by lithium-ion batteries (LIBs), which enjoy significant advantages in both power density and cycleability. But LIBs are at risk of catastrophic thermal runaway \cite{Loveridge2018}, and it has been shown that the material supply chain for LIBs may become uncertain in the future\cite{Vaalma2018}. New battery technologies based on sustainable and abundant materials are needed to safely and effectively meet our expanding energy storage needs. 

Zinc-based batteries, particularly zinc-air batteries (ZABs)\cite{Pang2017,Xu2015,Li2014,Pei2014a}, stand out as one of the most promising and mature battery technologies to complement LIBs \cite{Parker2017,Pang2017,Fu2017,Li2017b,Liu2017d}. Zinc metal is abundant, cheap, non-toxic, and practically stable in aqueous electrolytes. As a divalent metal, \ce{Zn} electrodes can achieve a very large specific capacity (819.9 \si{\milli\ampere\hour\per\gram}, 5853.8 \si{\milli\ampere\hour\per\liter}). The most suitable application for rechargeable ZABs is stationary energy storage\cite{Posada2017,Khor2018,Amunategui2017}, but they have also been proposed for EV applications\cite{Cano2018,Hannan2017,Sherman2018} and flexible electronics\cite{Tan2017b,He2017}. Primary zinc-air button cells with alkaline electrolytes (\textit{e.g.} \ce{KOH}) are widely used in hearing aids, due to their high practical energy density (circa 1000 \si{\watt\hour\per\liter})~\cite{Stamm2017}. However, carbonates form when the alkaline electrolyte is exposed to \ce{CO2} in the air, which limits the lifetime of the cell to just a few months~\cite{Stamm2017}. 

Carbonate formation is minimized in the near-neutral pH domain, and aqueous electrolytes containing ammonium chloride (\ce{NH4Cl}) have been proposed as a possible alternative to extend ZAB lifetime~\cite{Jindra1973,ThomasGoh2014,Sumboja2016}. ZABs featuring \ce{NH4Cl} electrolytes have been found to precipitate chloride-containing solids like \ce{Zn(NH3)2Cl2} and \ce{Zn5(OH)8Cl2*H2O}~\cite{Clark2019TowardsElectrolytes,ThomasGoh2014}, even though the final discharge product in a true zinc-air battery should be \ce{ZnO}. The precipitation of these mixed zinc salts consumes the electrolyte, passivates the electrode surfaces and lowers the energy density of the cell. Furthermore, because the pH stability in the air electrode requires an excess of buffering species, the slow diffusion of the weak acid or its conjugate base can become limiting at high current densities~\cite{Clark2017}. Electrolyte formulations that support a functionally stable pH and facilitate the precipitation of \ce{ZnO} are needed to improve the performance of near-neutral ZABs.


In this work, we highlight the complimentary and dynamic roles of weak acid dissociation, formation of complexes with dissolved \ce{Zn^{2+}} ions, and the solubility of zinc solids in stabilizing the electrolyte pH and facilitating the full discharge of the \ce{Zn}-air cell to \ce{ZnO}. By screening a variety of weak acids and counter-ions according to their acid dissociation constants ($\mathrm{K_a}$), \ce{Zn}-complex stability constants ($\mathrm{K_{eq}}$), and solubility product constants ($\mathrm{K_{sp}}$), promising electrolyte materials can be identified, simulated in thermodynamic and cell-level models, and experimentally validated. 

\begin{figure}[t]
  \includegraphics[width=1.05\columnwidth]{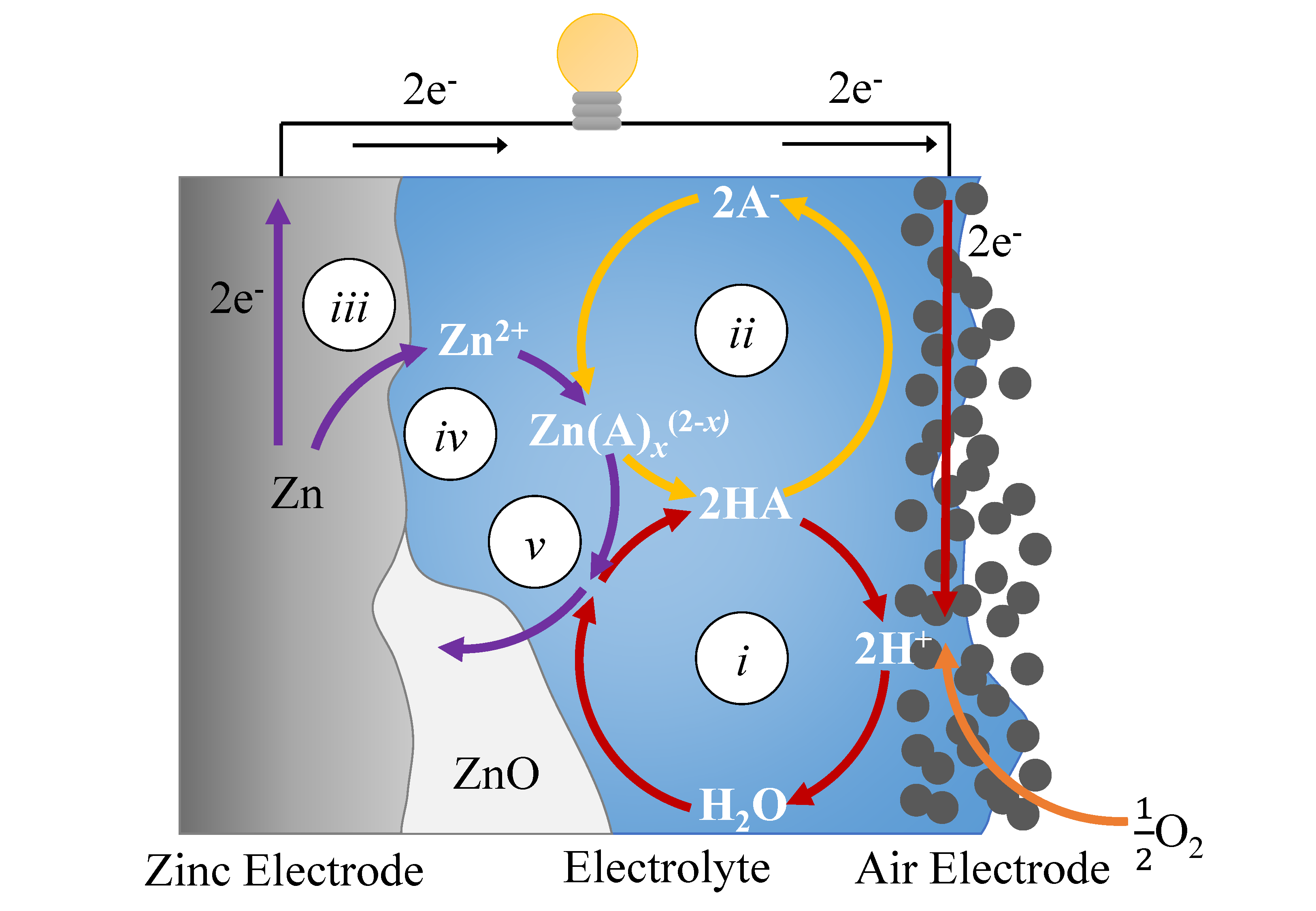}
  \caption{Schematic of idealized performance of ZABs with pH-buffered near-neutral electrolytes with some generic weak acid, \ce{HA}. Arrows indicate the direction during discharging. For a description, please refer to the text.}
  \label{fgr:Schematic}
\end{figure}

Fig. \ref{fgr:Schematic} shows the idealized operating principles of aqueous ZABs with pH-buffered near-neutral electrolytes. The main reactions are indicated by roman numerals \textit{i}-\textit{v} and are listed in the supplementary information$^\dag$. For descriptive purposes, we consider an electrolyte containing a generic weak acid, \ce{HA}. In the air electrode, dissolved \ce{O2} is reduced to form \ce{H2O} (reaction \textit{i}). The change in \ce{H+} concentration disturbs the equilibrium of the weak acid. The local loss of \ce{H+} causes the dissociation reaction, \ce{HA <=> H+ + A-}, to proceed to the right (reaction \textit{ii}). At the \ce{Zn} electrode, \ce{Zn^{2+}} is produced from the electrochemical oxidation reaction (reaction \textit{iii}) and forms complexes with other solutes in the electrolyte (reaction \textit{iv}), most importantly with the conjugate base of the weak acid, \ce{A-}. The formation of complexes between \ce{Zn^{2+}} and \ce{A-} enhances the pH stability of the electrolyte. When the solubility limit of zinc is exceeded, zinc solids precipitate (reaction \textit{v}). 

The selection of appropriate pH-buffering weak acid is open to some flexibility. Many inorganic pH buffers have acid dissociation constants and \ce{Zn}-complex formation constants in the appropriate range, but suffer from low solubility in electrolytes containing \ce{Zn^{2+}}. Organic weak acids (especially carboxylic acids) and their salts have been used for many years in small quantities as additives to concentrated alkaline electrolytes with the goal of improving Zn deposition and suppressing the corrosive hydrogen evolution reaction (HER). We propose that similar organic materials can fulfill a new role promoting pH stability and \ce{ZnO} precipitation. 


The remainder of this paper discusses the design of aqueous electrolytes to specifically address the challenges of near-neutral \ce{Zn}-air batteries. We begin by noting the design goals of the electrolyte and discussing the advantages and disadvantages of typical electrolyte components. After screening some classes of organic weak acids according to their dissociation and \ce{Zn}-complex stability constants, we perform thermodynamic calculations to predict their speciation and solubility characteristics across the pH domain. Having identified one possible promising electrolyte mixture containing citrate and glycine, the dynamic performance of the electrolyte is evaluated using simulated cell cycling. These simulations confirm the stability of the pH and give insight into the distribution of organic species during cell operation. The performance of the proposed electrolyte is experimentally validated using rotating ring disk electrode (RRDE) measurements to characterize the properties of different catalyst materials. Full-cell cycling measurements combined with \textit{operando} pH measurements confirm the stability of the electrolyte pH during discharging and charging. Additionally, ex-situ x-ray diffraction (XRD), scanning electron microscopy (SEM), and energy dispersive x-ray spectroscopy (EDS) characterization of the \ce{Zn} electrode confirm that \ce{ZnO} is present in the final discharge product.

\begin{table}[t]
\small
  \caption{\ Common zinc-based battery systems and aqueous electrolytes.}
  \label{tbl:ZnBatteries}
  \begin{tabular*}{\columnwidth}{@{\extracolsep{\fill}}ccc}
    \hline
    Systems & Aqueous electrolytes & References \\
    \hline
    \ce{Zn}-Air & \ce{KOH}, \ce{NH4Cl}, \ce{ZnCl2} & \citenum{Mainar2016a,ThomasGoh2014,Sumboja2016} \\
    \ce{Zn}-\ce{MnO2} &  \ce{KOH}, \ce{NH4Cl}, \ce{ZnCl2} & \citenum{Larcin1997,Zhao1998,Linden2004}  \\
    \ce{Zn} Redox Flow & \ce{ZnBr2}, \ce{ZnI2} & \citenum{Biswas2017,Zhang2018d,Xie2018}  \\
    \ce{Ni}-\ce{Zn} & \ce{KOH} & \citenum{Linden2004,Parker2017}  \\
    \ce{Zn}-Ion & \ce{ZnSO4}, \ce{Zn(CF3SO3)2} & \citenum{Kasiri2019MixedBatteries,Liu2019CalciumBatteries,Zhang2016Cation-DeficientBattery,Huang2019RecentBatteries} \\
    \hline
  \end{tabular*}
\end{table}

\section{Electrolyte design method}

To support ZAB operation, the electrolyte must fulfill a variety of basic requirements that include being ionically conductive with a moderate viscosity, high \ce{O2} solubility, and low vapor pressure. These requirements are met to some extent by many aqueous electrolyte materials and are described in more detail in the supplementary information$^\dag$. Near-neutral ZABs specifically require electrolyte materials that also provide a functionally stable pH and facilitate the precipitation of \ce{ZnO} as the final discharge product. In this work, we propose that achieving these goals depends on the selection of appropriate materials considering their acid dissociation constants, \ce{Zn}-complex formation constants, and solubility product constants. 

The selection of appropriate materials must strike a balance between several competing factors. For ZAB applications, we avoid very acidic pH values due to the very high zinc solubility and elevated risk of hydrogen evolution. On the other hand, we avoid very alkaline pH values due to the formation of carbonates from \ce{CO2}. Therefore, we focus on pH-buffered electrolytes with nominal pH values between circa 4 and 12. For the purpose of the discussion, these are referred to as near-neutral electrolytes (NNEs). The solubility of Zn solids in the near-neutral pH domain is typically very low, but can be improved through the presence of \ce{Zn^{2+}} complexing agents. The equilibrium formation constants of the resulting Zn-complexes should be selected such that they are high enough to boost the solubility of zinc in the electrolyte but low enough to still allow \ce{ZnO} precipitation and avoid unwanted \ce{Zn} electrode shape change. Next, we discuss what electrolyte materials are available to help achieve these goals.

\subsection{Electrolyte components}

A pH-buffered NNE requires two fundamental components: (i) a weak acid with at least one acid dissociation constant in the desired range whose conjugate base forms complexes with \ce{Zn^{2+}}, and (ii) a suitable counter-ion to maintain charge neutrality. Importantly, the counter-ion should not form insoluble products with \ce{Zn^{2+}}. 

Ammonium, \ce{NH4+}, is one of the most common pH buffers and requires a negatively charged counter-ion, often a halide like \ce{Cl-}, \ce{Br-}, \ce{F-}, or \ce{I-}. The solubility of zinc-halide-hydroxide solids in the near-neutral pH regime is low~\cite{Limpo1993,Limpo1995,Larcin1997,Arizaga2012}, which suppresses \ce{ZnO} as the dominant discharge product~\cite{Clark2017,Clark2019TowardsElectrolytes}. Furthermore, halide anions in electrolyte solutions are known to corrode non-noble metals and poison \ce{Pt}/\ce{C} ORR catalyst materials~\cite{Schmidt2001TheAnions,Malko2015TheCatalysts}, and elemental halogens, especially \ce{Cl2} and \ce{F2}, are toxic. Avoiding the use of halide counter-ions in the electrolyte could improve sustainability, performance, and lifetime of the battery. Aqueous zinc electrolytes with alternative negative counter-ions like \ce{SO4^{2-}}, \ce{NO3-}, \ce{ClO4-}, and others have been demonstrated~\cite{Jindra1973} but suffer from similar precipitation challenges (Fig. S2$^\dag$). Recent research has been directed to bulky anions like triflate, \ce{CF3SO3-},~\cite{Liu2019CalciumBatteries,Zhang2016Cation-DeficientBattery} and bistriflimide (\ce{TFSI}), \ce{(CF3SO2)2N-},~\cite{Wang2018a} but they are currently too expensive. It is difficult to identify a suitable negatively-charged counter-ion that is soluble, non-toxic, affordable, and stable under ZAB conditions. We therefore direct our attention towards alternative pH-buffers that can be utilized with positive alkali metal counter-ions (\textit{e.g.} \ce{Li+}, \ce{Na+}, and \ce{K+}).

\begin{figure*}[t]
  \includegraphics[width=1.0\textwidth]{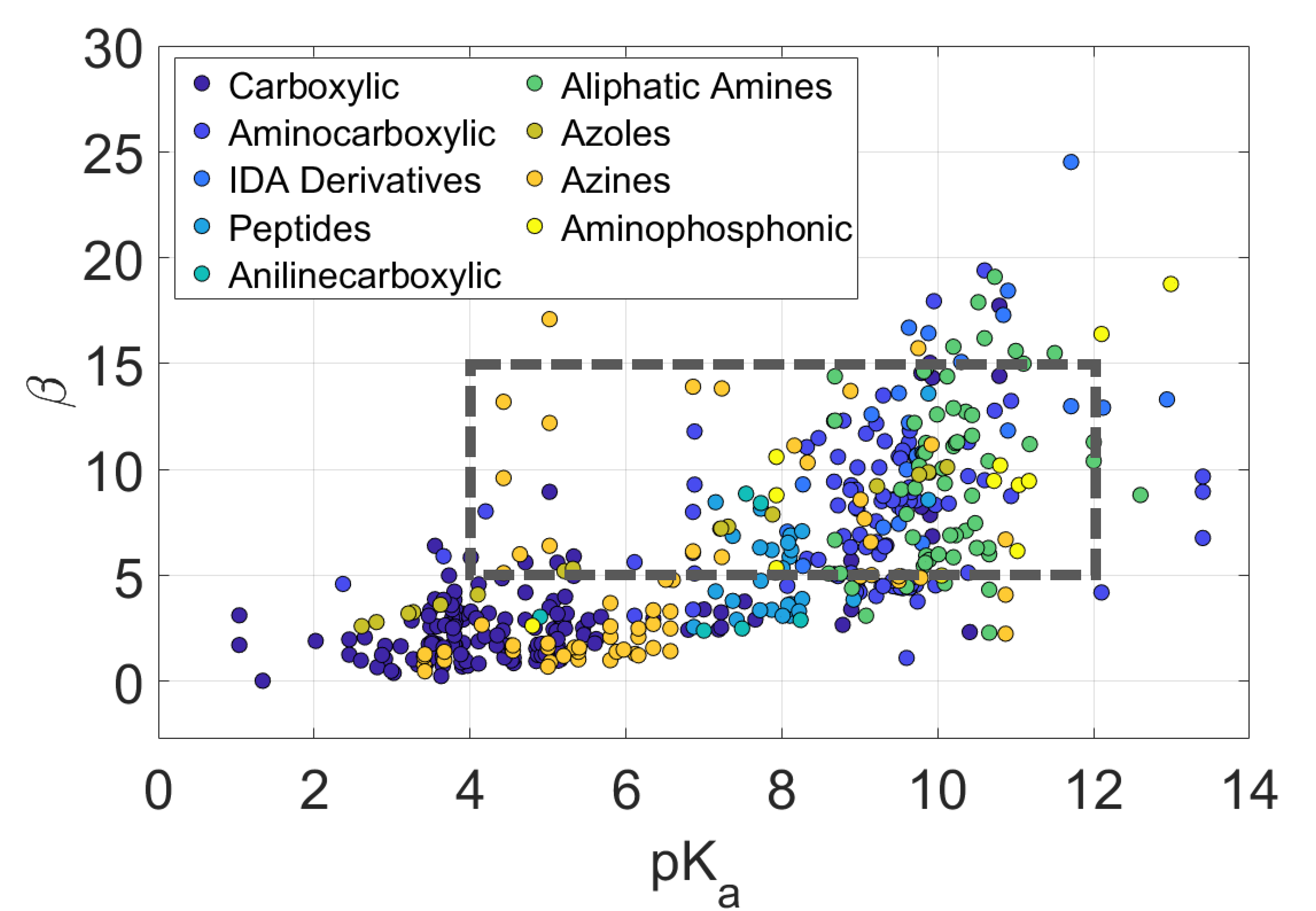}
  \caption{Zinc complex stability constant $\beta = \mathrm{log}_{10}K_{\mathrm{eq}}$ and $\mathrm{pK_a}$ values for a selection of weak organic acids. Dashed lines indicate the approximate region of interest for aqueous pH-buffered zinc electrolyte applications. Data compiled from Refs.~\citenum{Smith1976,Martell1977,Acids1977}}
  \label{fgr:OrganicData}
\end{figure*}

Organic weak acids, like carboxylic or aminocarboxylic acids, and their salts have been studied for many years as additives to aqueous zinc electrolytes to improve the quality of \ce{Zn} deposition and suppress the HER~\cite{Li2016TowardsNa3V2PO43,Ghazvini2018, Lei2018,Zhang2018d,Li2013,Lee2016b,Winand2011,Mclarnon1991a,Chen2015a,Liu2016,Huang2017,Mno2014,Xie2016a,Amend,Xie2017,Thomas1981,Ortiz-Aparicio2007,Ballesteros2011a}. These acids are often polyprotic, the conjugate bases are negatively-charged, and they can be combined with a positive counter-ion. The solubilities of zinc-organic salts are usually high enough so as not to threaten the precipitation of \ce{ZnO}. The wide variety of organic weak acids gives added flexibility in electrolyte design. Disadvantages of organic acids and their salts are the lower ionic conductivity and complicated redox characteristics, due to the variety of possible intermediate products. Nonetheless, they deserve closer inspection to determine their feasibility as aqueous zinc electrolyte materials. In the following section, we identify thermodynamic descriptors to screen suitable organic molecules and model solutions in chemical equilibrium. 

\subsection{Thermodynamic screening of organic components for aqueous electrolytes}

\begin{figure*}[t]
  \includegraphics[width=1.0\textwidth]{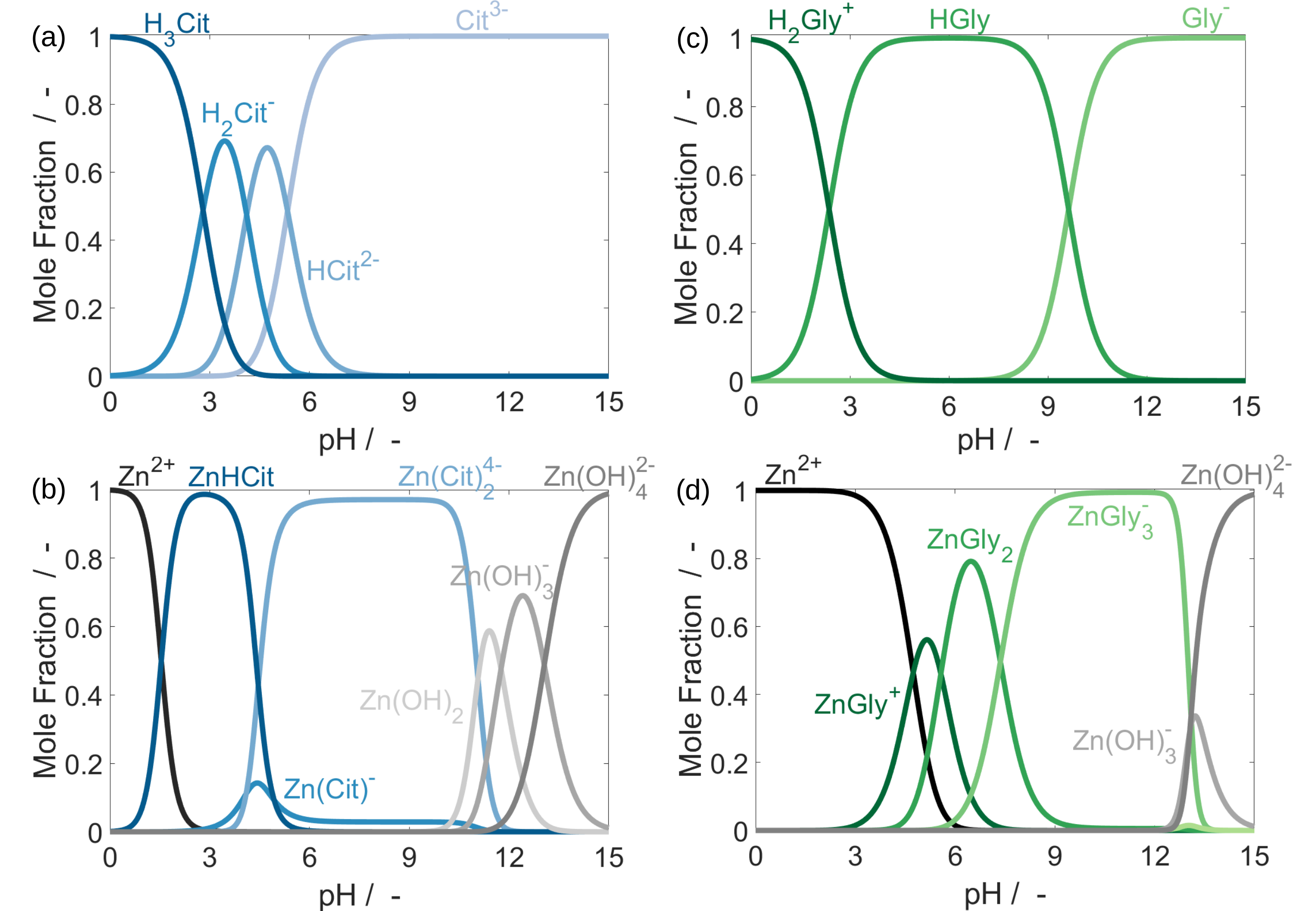}
  \caption{Themodynamic speciation and solubility plots for aqueous solutions of citric acid and glycine. Dissociation diagrams for 1M solutions (a), (c), and with 0.1 M $[\ce{Zn}]_{\mathrm{T}}$ added (b), (d).} 
  \label{fgr:OrganicSpeciation}
\end{figure*}

There are myriad organic molecules that could be suitable for aqueous zinc electrolytes. Two thermodynamic descriptors are selected to aid the material screening: $\mathrm{pK_a}$ values for dissociation of the weak acid and stability constants between the conjugate base and \ce{Zn^{2+}}. This data is compiled in the thorough work of Smith and Martell~\cite{Smith1976,Martell1977,Acids1977}. The modelling framework for the equilibrium speciation and solubility model applied in this section is derived in existing works~\cite{Limpo1993,Limpo1995,Clark2017,Clark2019TowardsElectrolytes} and described in the supplementary information$^\dag$.

Fig. \ref{fgr:OrganicData} presents a visualization of $\mathrm{pK_a}$ and logarithmic zinc complex stability constant ($\beta = \mathrm{log}_{10}K_{\mathrm{eq}}$) data for a variety of organic weak acids. There are many molecules that could be considered as pH buffers. To focus our search, we look for weak acids with one or more $\mathrm{pK_a}$ values in the near-neutral pH range, whose conjugate bases form moderately strong complexes with \ce{Zn^{2+}} and have electrochemical heritage. Among the more promising candidates are carboxylic acids and aminocarboxylic acids. Although both acetate and citrate have strong electrochemical heritage, we select citrate for further investigation in this analysis because of its superior pH buffering and transport properties~\cite{Newman,Apelblat1991}. The simplest aminocarboxylic acid is glycine, which has electrochemical heritage in battery\cite{Xie2017} and electroplating electrolytes\cite{Thomas1981,Ortiz-Aparicio2007,Ballesteros2011a}. 

Figs. \ref{fgr:OrganicSpeciation}(a) \& (b) show the dissociation and \ce{Zn^{2+}} speciation properties of an aqueous citric acid solution. In Fig. \ref{fgr:OrganicSpeciation}(a), citric acid (\ce{H3Cit}) dominates the solution at acidic values. As the pH increases, it passes through its various deprotonated states until citrate (\ce{Cit^{3-}}) is the sole species at pH values around 7 and above. When \ce{Zn^{2+}} ions are introduced to the solution in Fig. \ref{fgr:OrganicSpeciation}(b), they form complexes with the various citrate species in the solution, dominated by \ce{Zn(Cit)_2^{4-}} in the near-neutral pH domain. As the solution becomes more alkaline, the zinc-hydroxide complexes become dominant. This model of chemical equilibrium shows that pH-adjusted solutions of citric acid and zinc could stabilize the electrolyte pH between values of circa 3-6 due to citric acid dissociation and 11-14 due to the formation of zinc-hydroxide complexes. We do not work with these pH ranges because hydrogen evolution would limit efficiency at pH 3 and carbonate formation would limit lifetime at a pH 14.

Fig. \ref{fgr:OrganicSpeciation}(c) \& (d) present the dissociation and \ce{Zn^{2+}} speciation properties of glycine. Fig. \ref{fgr:OrganicSpeciation}(c) shows that, in aqueous solutions, glycine can exist in three states: the glycinium cation (\ce{H2Gly+}), the glycine zwitterion (\ce{HGly}), and the glycinate anion (\ce{Gly-}). When zinc is introduced in Fig. \ref{fgr:OrganicSpeciation}(d), it forms complexes with glycinate and the solution is dominated by \ce{Zn(Gly)_3^-} between pH 9-12. The best option to obtain a stable pH under ZAB operating conditions to utilize the \ce{HGly}/\ce{Gly-} buffer between pH 8-12, which coincides with the domain of \ce{Zn(Gly)_3^-} dominance and abuts the region of \ce{Zn(OH)4^{2-}} dominance. Glycine has good pH buffering properties in the appropriate domain and supports the controlled deposition of \ce{Zn} metal. Solutions of citric acid and its salts have high ionic conductivity and suppress \ce{H2} evolution on \ce{Zn} metal. Combining glycine with citric acid salt could further stabilize the pH - especially during ZAB charging - and improve charge transport within the electrolyte.

\begin{figure}[t]
  \includegraphics[width=\columnwidth]{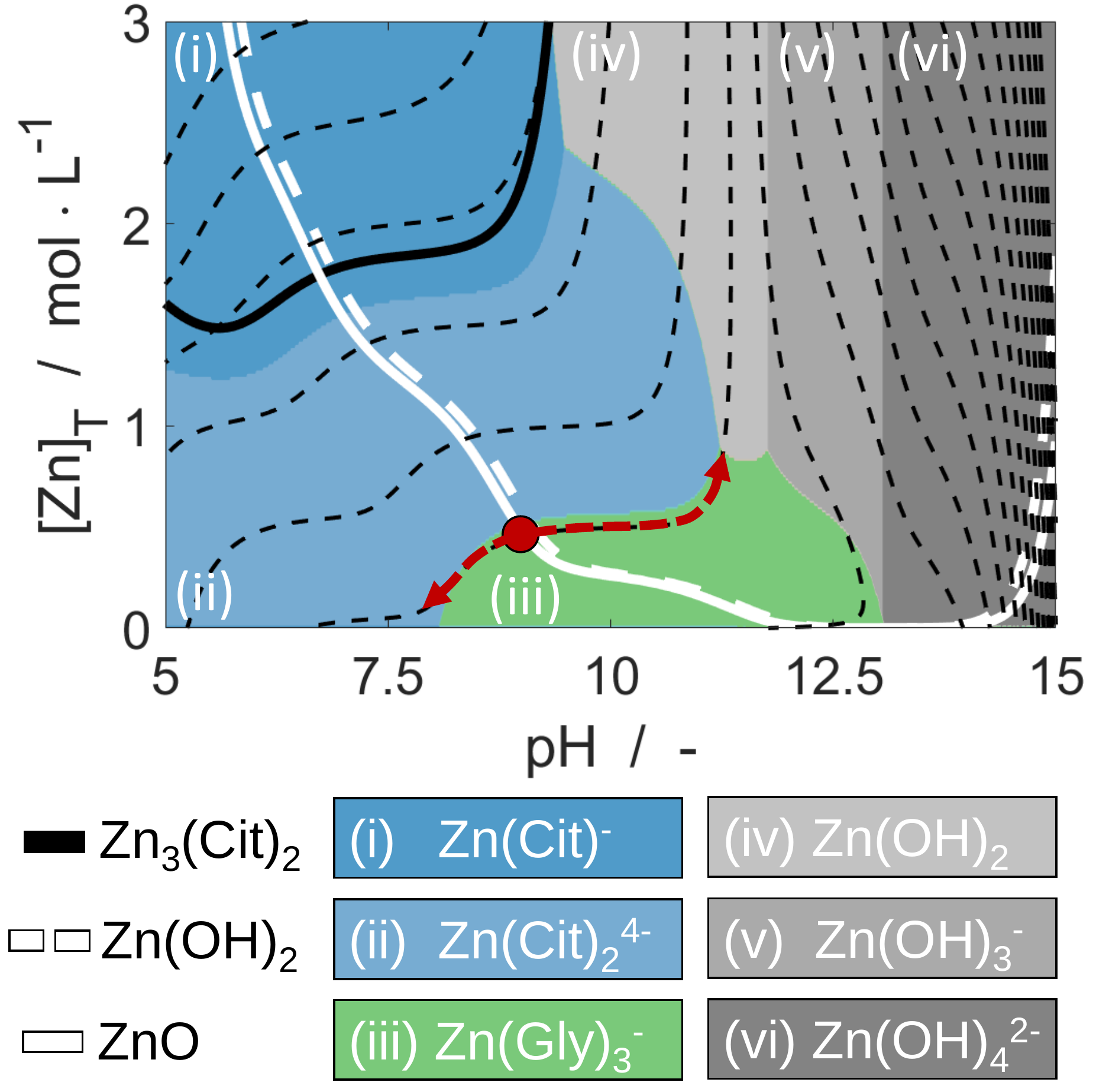}
  \caption{Equilibrium speciation and solubility landscape of the proposed citric acid - glycine electrolyte mixture, for fixed total concentrations $[\ce{Cit^{3-}}]_\mathrm{T}$ = 1.8 M and $[\ce{Gly^{-}}]_\mathrm{T}$ = 0.9 M. Colored regions (i) - (vi) indicate the dominat zinc complex in the electrolyte. Thick lines show the solubility limits of \ce{ZnO}, \ce{Zn(OH)2}, and \ce{Zn2(Cit)3}. The thin dotted lines trace paths of constant total \ce{K+} concentration. The super-imposed red dot indicates the proposed stable working point of the electrolyte, and the red arrows show the anticipated composition shifts during cell discharging and charging.}
  \label{fgr:CitGlyLandscape}
\end{figure}

Fig. \ref{fgr:CitGlyLandscape} shows the 2D solubility and speciation landscape of a mixed citric acid-glycine electrolyte with the pH adjusted through the addition of \ce{KOH}. The total concentration of citrate in the solution is 1.8 \si{\mole\per\cubic\deci\meter} ($[\ce{Cit^{3-}}]_\mathrm{T} = 1.8 \mathrm{M}$) and glycine is 0.9 \si{\mole\per\cubic\deci\meter} ($[\ce{Gly^{-}}]_\mathrm{T} = 0.9 \mathrm{M}$). The figure can be read as follows: the colored regions (labelled (i) - (vi)) represent the dominant zinc complexes in the electrolyte. The thick white lines show the solubility limits of \ce{ZnO} (solid) and \ce{Zn(OH)2} (dashed). The thick black line shows the solubility limit of \ce{Zn3(Cit)2}. The dashed black lines trace paths of constant [\ce{K+}] concentration, indicating how the composition of the electrolyte shifts as the cell is discharged ($[\ce{Zn}]_{\mathrm{T}}$ increases) or charged ($[\ce{Zn}]_{\mathrm{T}}$ decreases). Stable working points for the ZAB are located at positions where the dashed black lines (iso-[\ce{K+}] paths) cross the solid white line (\ce{ZnO} solubility). Locating the position on the chart corresponding to a total zinc concentration of 0.5 M and pH of 9, the dotted pathway shows stable operation between pH values of 8 and 11, with mixed \ce{ZnO}-\ce{Zn(OH)2} dominating the discharge product. There is a risk of zinc citrate precipitation, but this can be avoided with proper electrolyte preparation. Therefore, we propose an electrolyte containing 1.8M \ce{Cit^{3-}}, 0.9M \ce{HGly}, saturated with \ce{ZnO} and adjusted to pH 9 through the addition of \ce{KOH}.

There are a few aspects motivating our selection of the proposed halide-free electrolyte for further investigation. To our knowledge, this electrolyte mixture has not been previously proposed or investigated; it is therefore a suitable proof-of-concept for our electrolyte design rationale. Second, the dynamic behaviour of this electrolyte is very complicated and serves as an excellent opportunity to further develop and validate the proposed continuum modelling framework. Finally, these materials are very safe and cheap, and these ZABs could perhaps serve some nich\`{e} applications. Therefore, this selection should be seen as one step in the process of aqueous electrolyte development. In the following section, we discuss a method to simulate the cell-level dynamic performance of this aqueous organic electrolyte.

\section{Results and discussion}

The performance of a lab-scale ZAB with an aqueous organic electrolyte is investigated using both experimental measurements and cell-level simulations. In Sec. \ref{sec:simulations}, we begin by simulating the cell performance under galvanostatic cycling conditions to determine the feasibility of the proposed electrolyte and provide a foundation for understanding the experimental results. We then characterize the cell electrochemically with full cell and half cell measurements in Sec. \ref{sec.echemchar}, and characterize the cycled \ce{Zn} electrode physically to confirm the composition of the precipitate phase with XRD, SEM, and EDS measurements in Sec. \ref{sec:precipitation}. 

\subsection{Cell simulations}
\label{sec:simulations}
\begin{figure}[t]
  \includegraphics[width=\columnwidth]{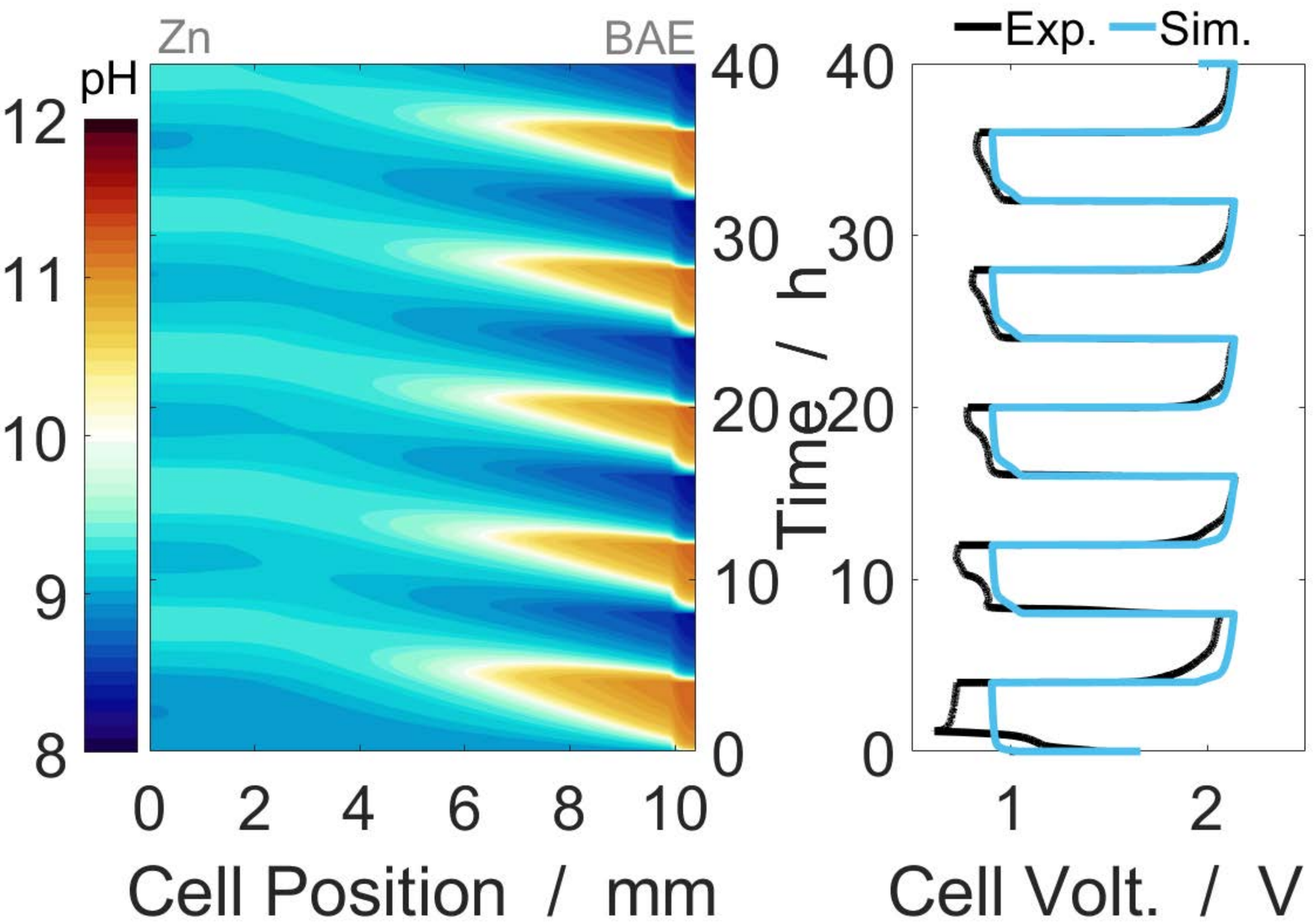}
  \caption{Simulated galvanostatic cycling performance of a lab-scale ZAB with an aqueous organic electrolyte, showing (a) electrolyte pH and (b) cell voltage. The cell is first discharged at 0.5 \si{\milli\ampere\per\square\centi\meter} for 4 \si{\hour} and then charged at 0.5 \si{\milli\ampere\per\square\centi\meter} for 4 \si{\hour} (see text for further details).}
  \label{fgr:OrganicpH}
\end{figure}

\begin{figure*}[t]
  \includegraphics[width=1.0\textwidth]{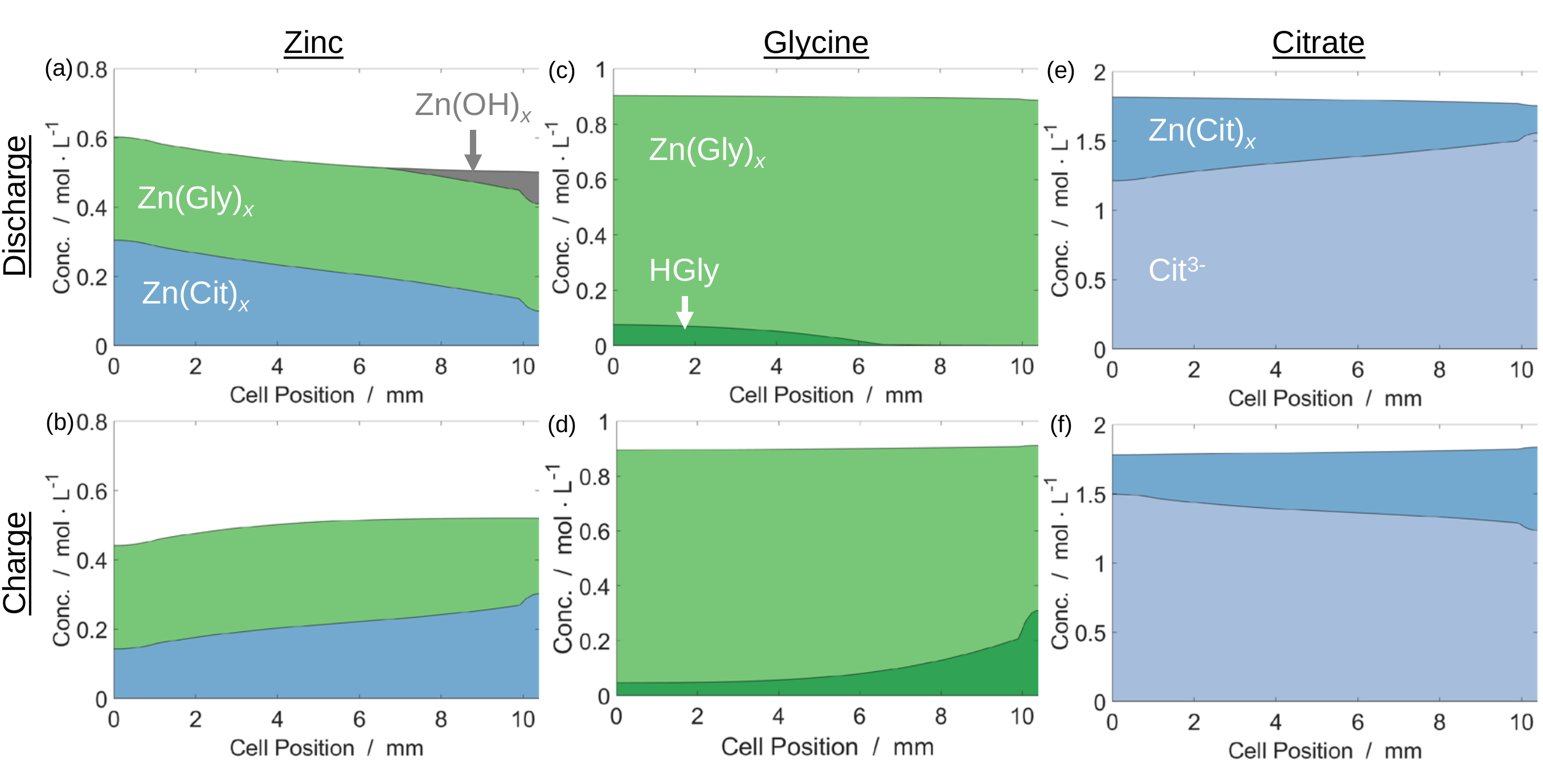}
  \caption{Anatomized concentration profiles of zinc (a)-(b), glycine (c)-(d), and citrate (e)-(f) in the electrolyte at the end of discharging and charging. The presence of \ce{Zn(OH)_{\textit{x}}} in the discharged electrolyte but not in the charged electrolyte, combined with the increased presence of \ce{HGly} in the charged electrolyte demonstrates a shift in the dominant pH buffering mechanism between discharging and charging. }
  \label{fgr:OrganicConc}
\end{figure*}

Fig. \ref{fgr:OrganicpH} shows (a) the simulated electrolyte pH profile and (b) a comparison of the simulated and experimental (EMD) cell voltage during galvanostatic cycling. The cell is first discharged at 0.5 \si{\milli\ampere\per\square\centi\meter} for 4 \si{\hour} and then charged at 0.5 \si{\milli\ampere\per\square\centi\meter} for 4 \si{\hour}. The \ce{Zn} electrode is at the left of the domain, followed by an electrolyte bath 9 \si{\milli\meter} thick, and the bi-functional air electrode (BAE) is on the right of the domain. The initial pH of the electrolyte is 9. 

At the start of discharge, the electrolyte becomes more alkaline in the BAE because of the effects of the ORR. The buffering capacity of the electrolyte stabilizes the pH at values around 11. As discharge continues, the electrolyte in the separator steadily trends alkaline due to the diffusion of spent buffer solution away from the air electrode. Because the electrolyte bath is relatively large, it takes time for this diffusion front to reach the \ce{Zn} electrode. The pH in the \ce{Zn} electrode varies only slightly (between circa 9-10) during cycling. To better understand the behavior of the electrolyte, we examine the distribution on zinc, glycine, and citrate in the electrolyte.

Fig. \ref{fgr:OrganicConc} shows the anatomized concentration profiles of zinc, glycine, and citrate in the electrolyte at the end of the first discharge and at the end of the first charge. At the end of the first discharge (Fig. \ref{fgr:OrganicConc}(a)), electrolyte zinc species in the BAE exists mostly as complexes with \ce{Gly-} with some zinc-hydroxides present. At the end of the first charge (Fig. \ref{fgr:OrganicConc}(b)), there are no zinc-hydroxides present in the BAE. This indicates that the alkaline shift that occurs during discharge is stabilized by the uptake of \ce{OH-} by the \ce{Zn(OH)_{\textit{x}}} complexes. On the other hand, the acidic pH shift that occurs during charging is stabilized by the \ce{HGly}/\ce{Gly-} buffer. 

This effect is also seen in the glycine distributions shown in Figures \ref{fgr:OrganicConc}(c) and (d). At the end of discharging, glycine in the air electrode exists only as complexes with zinc. But at the end of charging there is a significant increase in the proportion of glycine in its \ce{HGly} zwitterionic state. The concentration profiles of citrate (Figs. \ref{fgr:OrganicConc}(e) \& (f)) show that it mostly acts as a background electrolyte. Citrate does form some complexes with zinc, but the pH of the electrolyte does not drop to values low enough to engage its buffering properties. 

The cell-level simulations predict that a ZAB with the proposed citrate-glycine electrolyte can be cycled at low current densities. The pH is anticipated to stabilize between circa 8.5-11.5 in the air electrode and circa 9-10 in the \ce{Zn} electrode. In the following sections, lab-scale ZAB cells with the proposed electrolyte are experimentally characterized to investigate and validate these predictions.

\subsection{Physical and electrochemical characterization}
\label{sec.echemchar}

\begin{table*}[t!]
  \caption{Measured physicochemical properties of the proposed electrolyte composition compared with literature values for \ce{NH4Cl2-ZnCl2} and 30 wt\% \ce{KOH}. Values for ionic conductivity (IC), mass density ($\rho$), and dissolved oxygen concentration ([\ce{O2}]) are measured for each electrolyte. Values for \ce{NH4Cl-ZnCl2} are reported in ref.~\citenum{Clark2019TowardsElectrolytes}, and values for \ce{KOH} are reported in refs.~\citenum{Mainar2018a,Akerlof1941,Davis1967,Sipos2000}.}
  \label{tbl:ElectrolyteProperties}
  \def\arraystretch{1.5}
  \begin{tabular*}{\textwidth}{@{\extracolsep{\fill}}lccccccc}
    \hline
    Electrolyte  & pH & IC ($\mathrm{mS} \cdot \mathrm{cm}^{-1}$) & $\rho$ ($\mathrm{g} \cdot \mathrm{mL}^{-1}$) & [\ce{O2}] ($\mathrm{mg} \cdot \mathrm{L}^{-1}$) \\
    \hline
    1.8M \ce{K3Cit} - 0.9M \ce{HGly} & 12 & 93.3 & 1.3726 & 2.96 \\
    1.8M \ce{K3Cit} - 0.9M \ce{HGly} & 9 & 86.8 & 1.338 & 2.76 \\
    1.8M \ce{K3Cit} - 0.9M \ce{HGly} & 8 & 90.8 & 1.3483 & 2.81 \\
    1.6M \ce{NH4Cl} - 0.5M \ce{ZnCl2} & 8 & 209 & 1.05 & 6.61  \\
    30 wt\% \ce{KOH} & 14.8 & 638 & 1.28 & 2.52 \\
  \end{tabular*}
\end{table*}

Table \ref{tbl:ElectrolyteProperties} lists the properties of the as-prepared electrolyte compared with previously reported values for aqueous ZAB electrolytes \ce{KOH}~\cite{Mainar2018a,Akerlof1941,Davis1967,Sipos2000} and \ce{NH4Cl-ZnCl2}~\cite{Clark2019TowardsElectrolytes}. The properties of the proposed electrolyte are measured at three pH values: 8, 9, and 12. These reflect the electrolyte pH values during charging, open-circuit, and discharging predicted by the cell simulations. The electrolyte properties at these reaction conditions are found to be similar. The proposed electrolyte has lower ionic conductivity than other ZAB electrolytes. On the other hand, density and dissolved oxygen concentration results are similar to other aqueous electrolytes. A total organic carbon (TOC) comparison of the as-prepared and cycled electrolyte is given in Table S11$^\dag$.

Electrochemical measurements are performed to evaluate the feasibility of the proposed electrolyte. First, rotating ring disk electrode (RRDE) measurements are used to compare and contrast the catalytic performance of CNT, EMD, and EMD+CNT air electrode materials. We then compare the cycling performance of the different catalysts in full cells, and long-term cycling tests investigate the cycle lifetime.  Finally, \emph{operando} pH measurements probe the stability of the electrolyte composition during discharging and charging and the performance of the ZAB as a primary cell is shown over a single complete discharge.

\subsubsection{Oxygen electrocatalysis}

\begin{figure}[t]
  \includegraphics[width=\columnwidth]{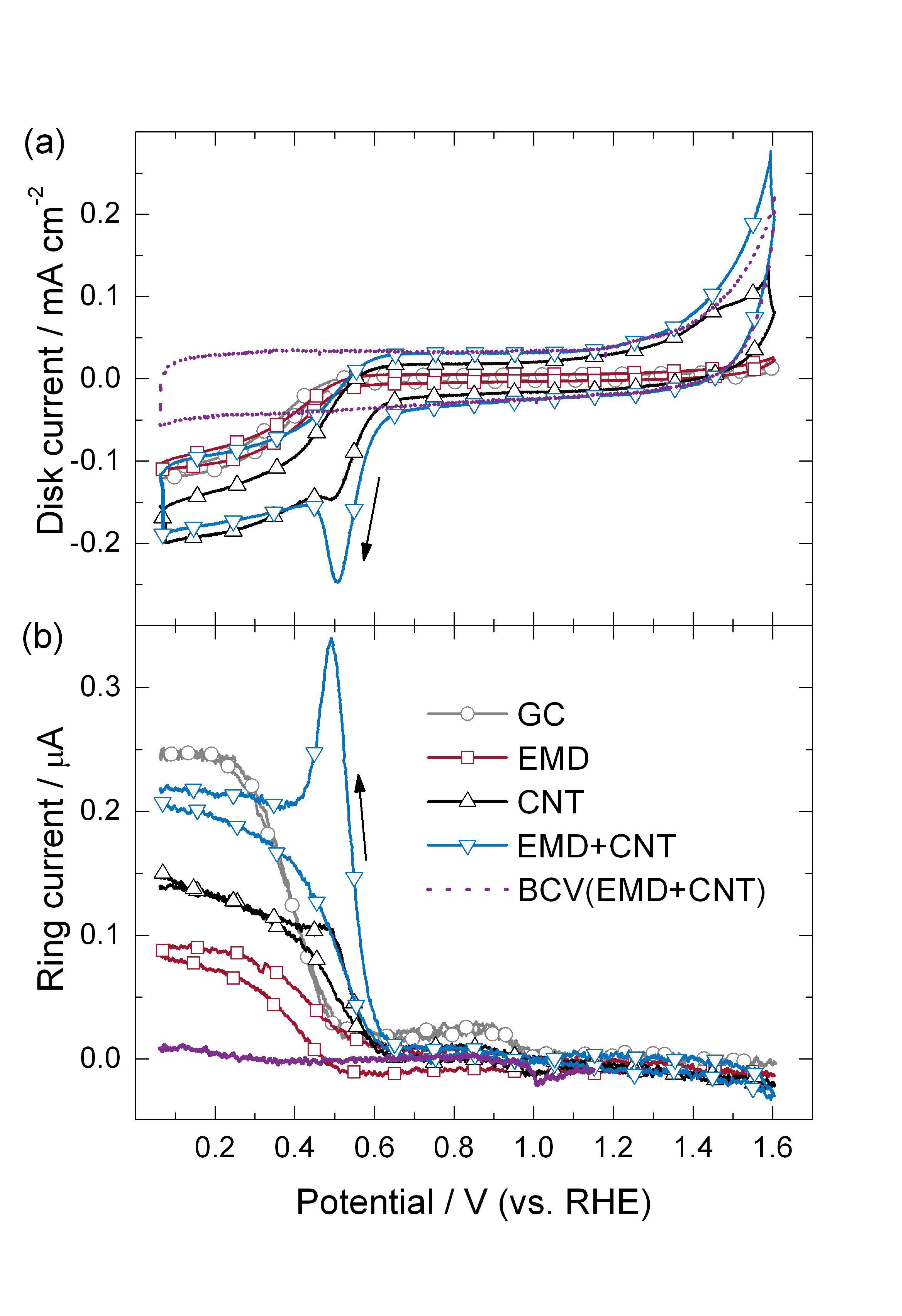}
  \caption{RRDE measurements at bare GC and thin-film EMD, CNT, and EMD+CNT electrodes, in oxygen-saturated (solid lines) and oxygen-free (dotted line) conditions. The electrolte is 1.8M \ce{K3Cit} - 0.9M \ce{HGly} saturated by \ce{ZnO} at pH 9. Currents are shown for (a) the disk electrode and (b) the ring electrode. $\omega$ = 1600 rpm (ORR), scan rate = 10 \si{\milli\volt\per\second}, $U_{\textrm{ring}}$ = 1.2 \si{\volt}. Shape markers are added to help distinguish the different measurements.}
  \label{fgr:RRDE}
\end{figure}

\begin{figure}[t]
  \includegraphics[width=\columnwidth]{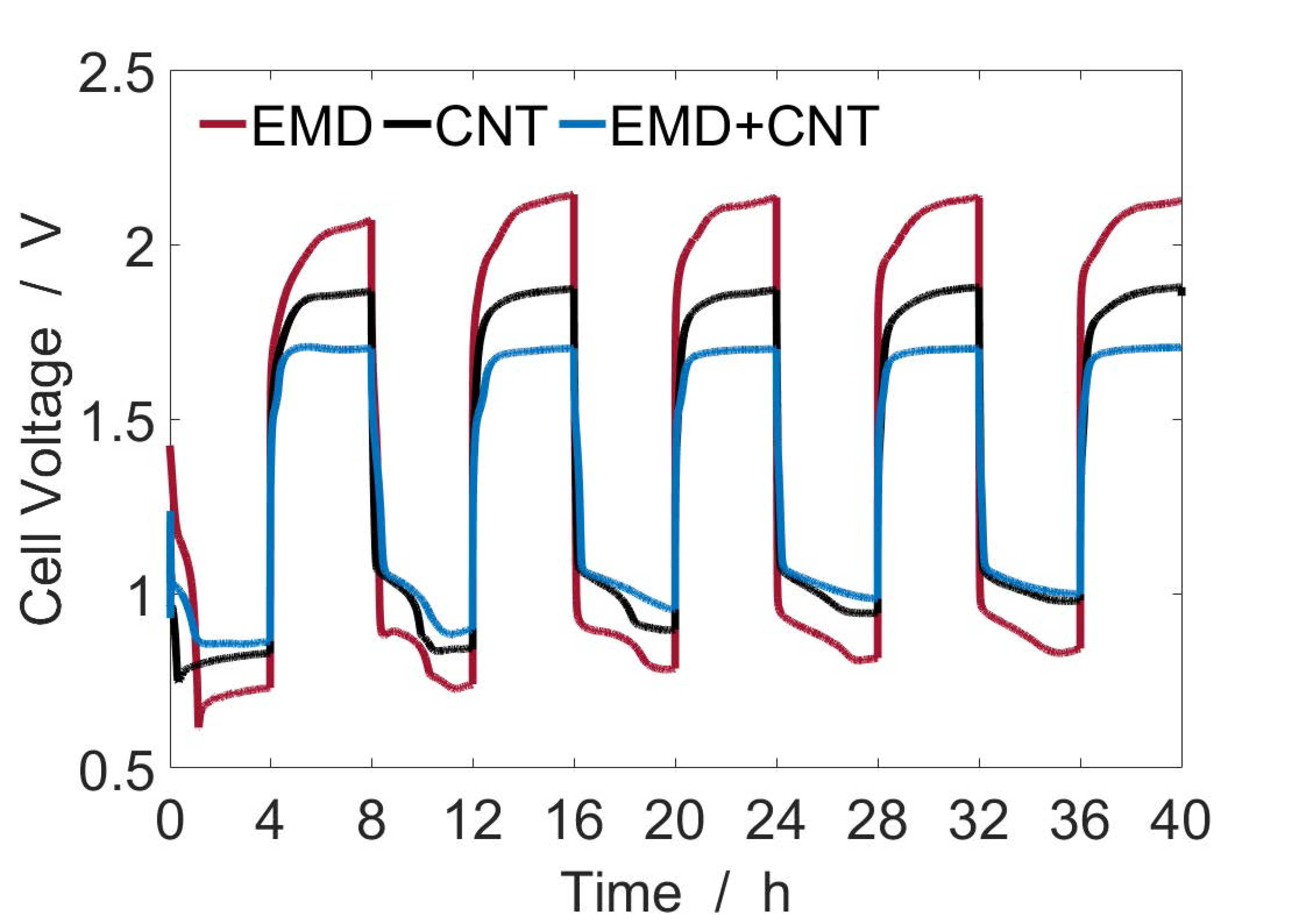}
   \caption{Cell voltage during galvanostatic discharge of a ZAB with CNT, EMD, and EMD+CNT catalyst materials over 5 cycles.}
  \label{fgr:Catalyst}
\end{figure}

The performance of air electrode catalyst materials in the proposed electrolyte are characterized using RRDE measurements. Reference measurements performed on the common polycrystalline Pt (pc-Pt) electrode in the proposed electrolyte without and with oxygen are presented in Fig. S4 of the supplementary information$^\dag$. 
Fig. \ref{fgr:RRDE} shows the cyclic voltammetry (CV) results for the various air electrode materials performed in the RRDE configuration. Since the glassy carbon (GC) disk was used as a substrate for the thin film electrode fabrication, comparative RRDE measurements in the proposed electrolyte were performed over a bare GC electrode and then covered by the individual components: EMD, CNT, and the EMD+CNT composite. 

Fig. \ref{fgr:RRDE}(a) shows the measured disk currents. Three common potential regions can be distinguished for all electrodes: (i) double-layer region from circa 0.7 \si{\volt} to circa 1.2 \si{\volt}; (ii) oxygen reduction reaction (ORR) region from circa 0.7 \si{\volt} to 0.07 \si{\volt}; and (iii) oxidation of organics and oxygen evolution reaction (OER) region from circa 1.2 \si{\volt} to 1.6 \si{\volt}. 

In the double-layer region, the pseudo-capacitive current corresponds to the electrochemically active surface area of the electrodes. The observed pseudo-capacitive current is the narrowest for the GC and EMD-covered GC electrodes, much wider for the CNT-covered GC electrode, and the widest for the EMD+CNT composite over the GC substrate. This demonstrates the greater active surface area of CNT materials.

The ORR sets in at circa 0.7 V for all the electrodes and reaches comparable ORR values at the lower potential limit, considering superimposed pseudo-capacitive contributions, even for the pc-Pt electrode (Fig. S4$^\dag$). This indicates that the ORR to proceed over a pc-Pt electrode largely blocked by adsorbed species, as supported by strongly suppressed underpotential hydrogen adsorption/desorption features of the base voltammetry in oxygen-free electrolyte$^\dag$. The CNT catalyst exhibits a shoulder in the ORR current at circa 0.5 \si{\volt} and the EMD+CNT catalyst shows an expressed peak at the same potential. A steeper onset in the ORR at the CNT-containing electrodes, compared to bare GC and EMD-covered GC can tentatively be explained by the storage of oxygen in the CNTs due to their tubular structure, which is supplied from the CNT reservoir in addition to the mass transport delivered oxygen during ORR onset during the negative-sweep scan to compensate the depletion of oxygen at the electrode due to the ORR. The appearance of this ORR shoulder in the negative-sweep scan is also in agreement with previous findings~\cite{Jiang2015,Kruusenberg2009}. Similar effects were also addressed regarding the behavior of porous rotating disk electrodes~\cite{Bonnecaze2007} and the effect of particle nano-morphology on the mass transport controlled reactions~\cite{Jiao2018}.

Ring currents are shown in Fig. \ref{fgr:RRDE}(b). The hydrogen peroxide (\ce{H2O2}) formation via the 2-electron ORR pathway at GC electrode increases at lower potentials and approaches stable values, which corresponds to circa 17\% \ce{H2O2} yields. The \ce{H2O2} formation at the CNT-film electrode is decreased by roughly half compared with a bare glassy carbon electrode, indicating an improved ORR selectivity at the rough electrode due to the increased probability for re-adsorption and further reduction of incomplete reaction intermediates~\cite{Westbroek2000,Seidel2008}. The ORR at the EMD+CNT composite electrode shows increased \ce{H2O2} formation at the peak of the ORR at circa 0.5 \si{\volt}, which indicates no change in the ORR selectivity.

The peroxide yield of circa 17\% during the ORR is comparable to other recent studies of bi-functional ORR/OER catalysts. In neutral media, \ce{Mn}, \ce{Fe}, \ce{Co} and \ce{Ni}-aminoantipyrine \cite{Rojas-Carbonell2017} or \ce{Cu}-phenantroline \cite{Yang2017} based catalysts show ORR selectivity close to four electrons. However, their OER activity and stability under the OER conditions are not yet reported, which is essential for the bi-functional performance of the catalyst. On the other hand, oxygen-doped carbon nanotubes are reported to be a highly selective catalyst for peroxide formation during the ORR in neutral media \cite{Lu2018}. To improve the ORR selectivity a further reduction of hydrogen peroxide can be ensured by engineering the loading, morphology and composition of the catalyst \cite{Jaouen2009, Poux2014,Ryabova2016}.        

Finally, the behavior at high potentials is attributable to a mix of organic species oxidation with contributions from the OER and carbon corrosion. In the base CV on pc-\ce{Pt}$^\dag$ (which is used as an active material for the oxidation of organic species and oxygen evolution), oxidation of the organic electrolyte species sets in at circa 1.2 \si{\volt}, as expected for a typical Kolbe-type reaction for decarboxylation of carboxylic acids \cite{Vijh1967} in both nitrogen and oxygen saturated electrolyte.

In Fig. \ref{fgr:RRDE}, the disk current for the GC substrate material used for the film electrode fabrication (grey) presents virtually no response until a slight increase above 1.5 \si{\volt}, which can be attributed to carbon corrosion. Similarly, the EMD catalyst material (blue) shows no response below 1.5 \si{\volt}, in agreement with low activity of electrodeposited manganese oxide towards the OER below 1.5 V in alkaline solution~\cite{Kolbach2017}. The current uptick above 1.5 \si{\volt} can be attributed to corrosion of the GC substrate with a possible contribution from the OER, which is expected to onset between 1.5 and 1.6 \si{\volt}~\cite{Yi2017}.

The CNT film electrode current (black) presents the onset of an oxidation process at circa 1.2 \si{\volt}, which aligns with the organic species oxidation process observed on pc-\ce{Pt} (Fig. S4$^\dag$). A small shoulder in the current is visible at 1.45 \si{\volt}. The EMD+CNT film current (red) traces the CNT current until circa 1.45 \si{\volt}. Above this potential, EMD+CNT shows close to an exponential increase in current in both nitrogen-saturated and oxygen-saturated electrolyte. The observed current growth is greater than the sum of the CNT and EMD individual contributions, pointing to a synergetic effect in the EMD+CNT mixture. The absence of a peak or plateau in the EMD+CNT disk current indicates that the process is not transport-limited. Furthermore, CNTs exhibit a high stability against carbon corrosion in this potential range~\cite{Mette2012}. An improved OER performance from the manganese oxide particles deposited onto nitrogen doped CNTs at potential higher that ca. 1.6 V was reported in ~\cite{Mette2012,C7CP02717F}, however, the EMD+CNT composite shows improved current growth above circa 1.5 \si{\volt}. Further investigations of this observation could be a topic for future research.

\begin{figure}[t]
  \includegraphics[width=\columnwidth]{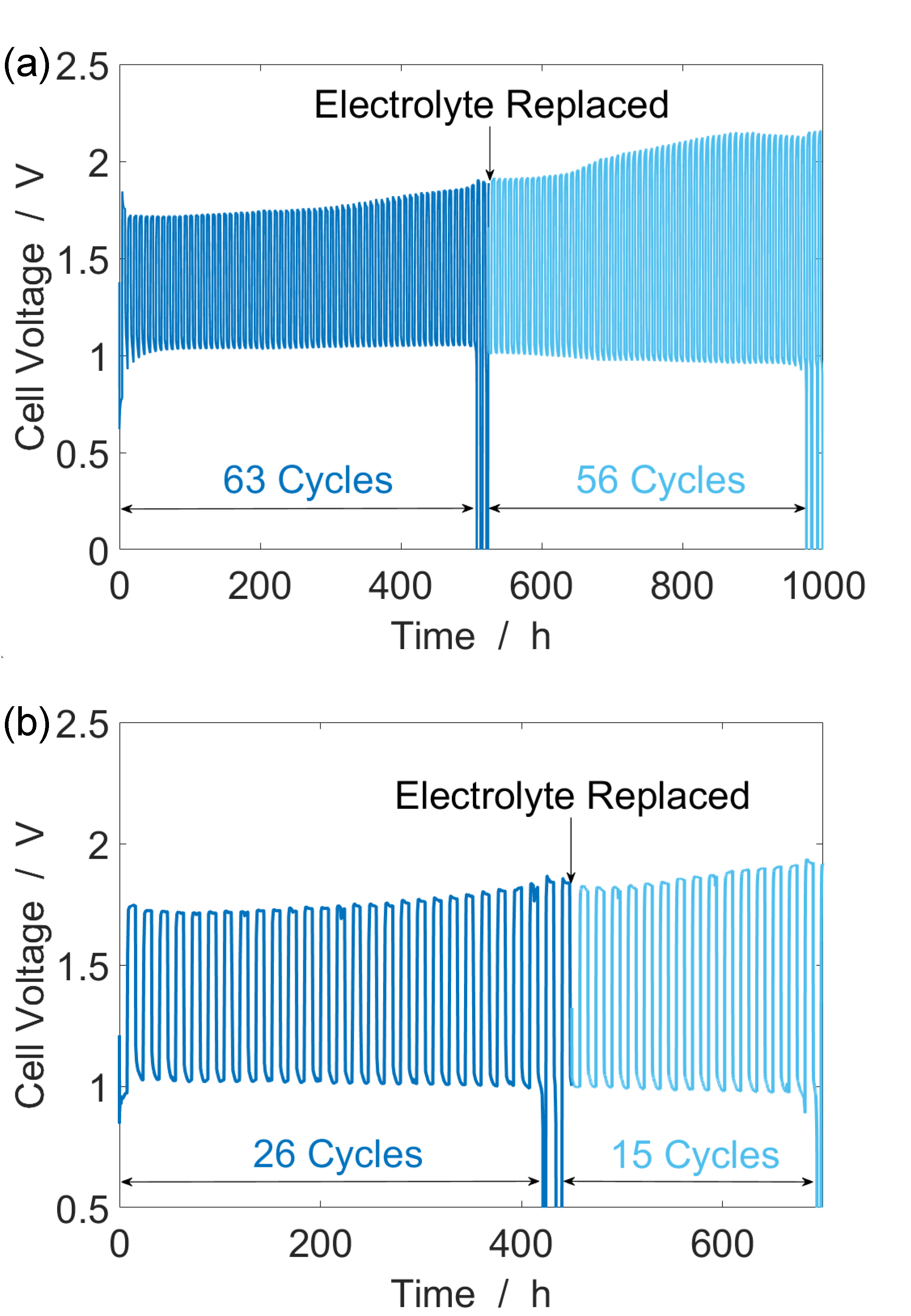}
   \caption{Cell voltage during long-term galvanostatic cycling at 0.5 \si{\milli\ampere\per\square\centi\meter} for (a) 4 \si{\hour} of discharging and 4 \si{\hour} of charging and (b) 8 \si{\hour} of discharging and 8 \si{\hour} of charging. The air electrode catalyst is EMD+CNT.}
  \label{fgr:Cycling}
\end{figure}

The main findings of the model RRDE studies are summarized as follows: (i) the comparison of different catalyst materials in nitrogen-saturated and oxygen-saturated electrolytes shows that each material is active towards the ORR, with EMD+CNT presenting the highest activity. Furthermore, significant ring currents are present in the low potential region, indicating the presence of \ce{H2O2} resulting from the 2-electron ORR pathway. (ii) GC and EMD show no response below 1.5 \si{\volt}, indicating that they are not active towards oxidation of the organic species. On the other hand, the oxidation of the organic species on CNT and EMD+CNT sets in between 1.2 \si{\volt} and 1.45 \si{\volt}. In the high potential range, the oxidation of organic components in the electrolyte is likely the dominant degradation mechanism. (iii) Above 1.45 \si{\volt} the EMD+CNT catalyst shows non-additive behavior compared to EMD and CNT alone. In this domain, EMD+CNT disk current exhibits non-transport-limited exponential growth. Such behavior can be attributed to a contribution from the OER in addition to oxidation of the organic species. Electrocatalytic decarboxylation via the Kolbe reaction mechanism \cite{Vijh1967} presents a significant challenge for the long-term electrochemical stability of both carboxylic and aminocarboxylic acids in rechargeable ZABs. The OOH* intermediate formation during the OER will inevitably affect the stability of both the electrolyte and the electrode material . One possible solution for further research could be to use other materials highlighted in Fig. \ref{fgr:OrganicData} that do not contain carboxyl groups, such as imidazole.

\begin{figure}[t]
  \includegraphics[width=\columnwidth]{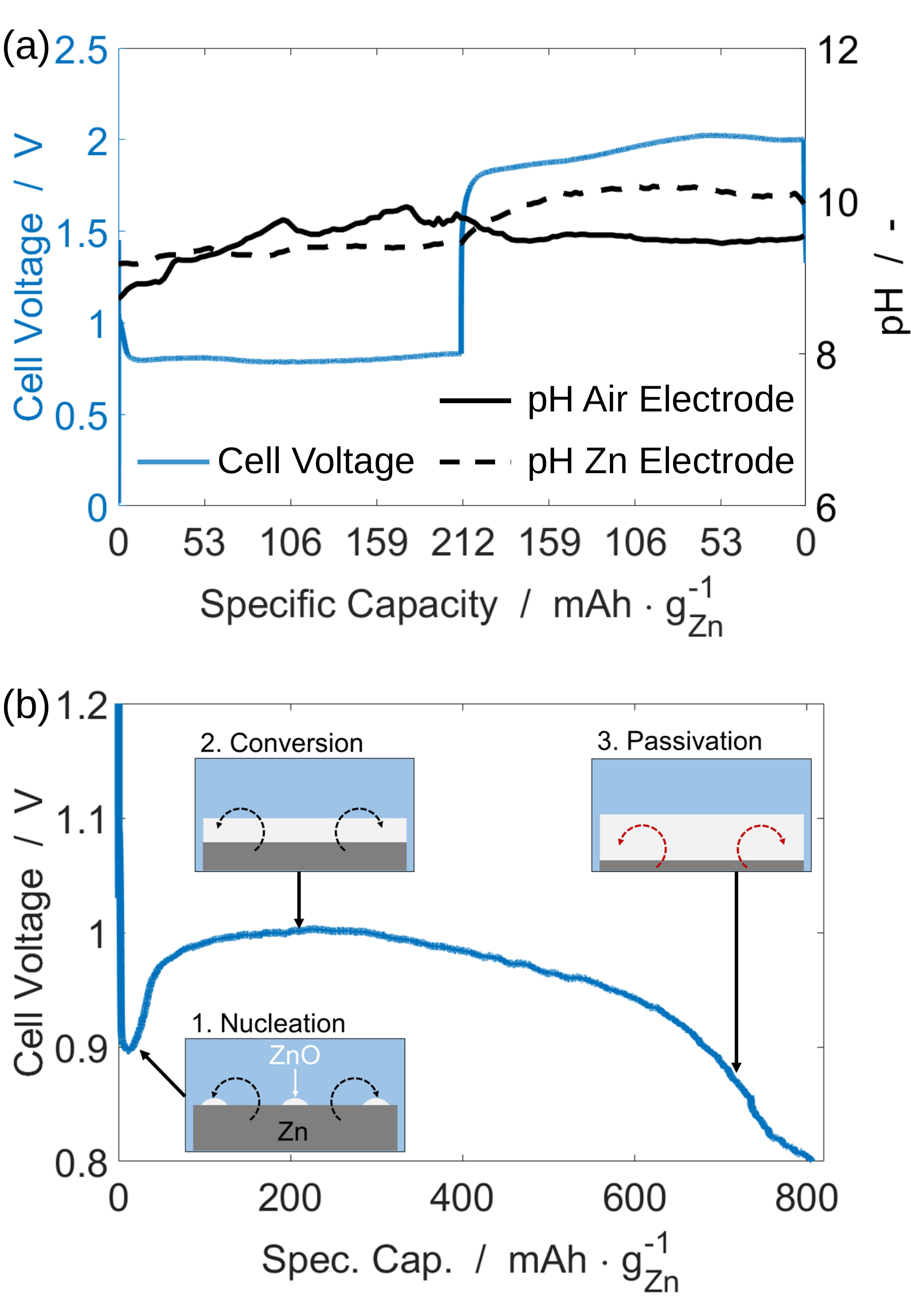}
   \caption{(a) Cell voltage and operando pH measurements near the air and \ce{Zn} electrodes during a single cycle at 2 \si{\milli\ampere\per\square\centi\meter} for 25 \si{\hour} of discharging and 25 \si{\hour} of charging. The air electrode catalyst is EMD+CNT. (b) Cell voltage during complete discharge at 0.5 \si{\milli\ampere\per\square\centi\meter} to cutoff voltage of 0.8 \si{\volt}. The air electrode catalyst is EMD+CNT. Discharging occurs in three stages. (1) The cell potential dips at the start of discharging due to nucleation of \ce{ZnO}. (2) As \ce{ZnO} continues to precipitate, the cell reaches its stable working point and the voltage levels off. (3) The cell voltage drops as the amount of usable \ce{Zn} metal is depleted. The cell ultimately achieves 805.7 $\mathrm{mAh} \cdot \mathrm{g}^{-1}_{\ce{Zn}}$, or 98.3\% of its theoretical capacity.}
  \label{fgr:Opp_pH}
\end{figure}

\subsubsection{Full cell measurements}

Fig. \ref{fgr:Catalyst} compares the cell voltage of a lab-scale ZAB cell with CNT, EMD, and EMD+CNT catalyst materials over 5 galvanostatic cycles. The cycling is performed at a current density of 0.5 \si{\milli\ampere\per\square\centi\meter} for 4 hours of discharging and 4 hours of charging (2 \si{\milli\ampere\hour\per\square\centi\meter} charge transferred per cycle). After 5 cycles, the discharging performance of ZABs with CNT and EMD+CNT catalyst materials converge to similar voltages (0.98 \si{\volt} and 1.00 \si{\volt}, respectively), while the discharging voltage of the ZAB with EMD remains at 0.85 \si{\volt}. On the other hand, ZAB cells with the different catalyst materials exhibit consistently different charging voltages. The cell with EMD catalyst shows a final charging voltage of 2.12 \si{\volt}, while the cell with CNT charges at 1.88 \si{\volt} and the cell with EMD+CNT at 1.72 \si{\volt}. This observation agrees with the catalyst trends shown in the half-cell RRDE measurements in Fig. \ref{fgr:RRDE}. The EMD+CNT catalyst was therefore selected as the preferred catalyst material for the remainder of the experimental characterization.

Continuous full cell cycling tests were performed with the EMD+CNT catalyst under various conditions to evaluate the long-term operation of the ZAB. Fig. \ref{fgr:Cycling} shows the cell voltage of ZAB cells continuously cycled at 0.5 \si{\milli\ampere\per\square\centi\meter}. In Fig. \ref{fgr:Cycling}(a), a cell cycled for 4 hours of discharging and 4 hours of charging demonstrates stable operation of 63 cycles over 504 hours (126 \si{\milli\ampere\hour\per\square\centi\meter} total charge transferred). After the first few cycles, the discharge voltage of the cell stabilizes at just over 1 \si{\volt}. The discharge voltage remains very stable throughout the cycling experiment, before undergoing a sudden drop after 63 cycles. On the other hand, the charging voltage slowly increases from  1.72 \si{\volt} at the beginning to 1.88 \si{\volt} by cycle 63. The electrolyte is then replaced and the cell recuperates some of its original performance. The discharging voltage stabilizes between 0.9 \si{\volt} and 1 \si{\volt}, but the charging voltage continues to increase to a peak of 2.15 \si{\volt}. After replacing the electrolyte the ZAB operates for an additional 56 cycles.

In Fig. \ref{fgr:Cycling}(b), the cycling time is increased to 8 hours of discharging and 8 hours of charging (4 \si{\milli\ampere\hour\per\square\centi\meter} charge transferred per cycle). The cell demonstrates 26 cycles over 416 hours (104 \si{\milli\ampere\hour\per\square\centi\meter} total charge transferred). Once again, the discharge voltage is stable around 1 \si{\volt}, but the charging voltage slowly increases with cycle number. When the electrolyte is replaced, the cell recovers and is able to operate for another 15 cycles. The cell cycled at 4 \si{\milli\ampere\hour\per\square\centi\meter} per cycle (Fig. \ref{fgr:Cycling}(b)) fails sooner than the cell cycled at 2 \si{\milli\ampere\hour\per\square\centi\meter} per cycle (Fig. \ref{fgr:Cycling}(a)), as quantified in Table S10. Replacing the electrolyte does not result in a complete recovery of cell performance, which indicates that some degradation of the electrodes is likely occurring.

The galvanostatic cell cycling performance indicates that electrolyte oxidation during charging is likely the dominant degradation mechanism in the cell. The RRDE measurement of EMD+CNT in Fig. \ref{fgr:RRDE} suggests that the OER could contribute a larger fraction of the current at higher potentials. ZAB cells charged with a constant voltage protocol demonstrated a modest gain in cycle lifetime, as shown in Fig. S6$^\dag$. 

The results of the continuous cycling tests confirm that the lab-scale ZAB cell with the proposed aqueous organic electrolyte can be cycled over an extended period (500 hours, 63 cycles). However, the observed increase in charging voltage and the recovery of performance after replacing the electrolyte suggest that some electrolyte degradation occurs during the charging process. The effects of this degradation can be lessened by applying a constant voltage charging protocol. An overview of all the results obtained during the various cycling tests is given in Table S10$^\dag$. 


In Fig. \ref{fgr:Opp_pH}(a), the stability of the electrolyte pH during a single discharge-charge cycle is investigated using \textit{operando} pH measurements near the \ce{Zn} and air electrodes. A ZAB cell with the proposed electrolyte cycled at 2 \si{\milli\ampere\per\square\centi\meter} for 25 hours of discharging and 25 hours of charging (50 \si{\milli\ampere\hour\per\square\centi\meter} charge transferred), the \textit{operando} measurements show that the electrolyte pH remains stable between circa 8.25 and 10.25. This is in agreement with the stable pH range predicted by both the thermodynamic model in Fig. \ref{fgr:CitGlyLandscape} and the dynamic cell-level simulations in Fig. \ref{fgr:OrganicpH}.

Finally, Fig. \ref{fgr:Opp_pH}(b) shows the cell voltage profile for one complete discharge of a ZAB with the proposed electrolyte at 0.5 \si{\milli\ampere\per\square\centi\meter}. The cutoff voltage is set as 0.8 \si{\volt}. At the end of discharging, the cell achieves a capacity of 805.7 $\mathrm{mAh} \cdot \mathrm{g}^{-1}_{\ce{Zn}}$, corresponding to 98.3\% utilization of the theoretical capacity of the \ce{Zn} electrode (819.9 $\mathrm{mAh} \cdot \mathrm{g}^{-1}_{\ce{Zn}}$). The measured ZAB discharge occurs in three stages. (1) At the start of discharge, there is an initial dip in cell voltage, which can be attributed to the nucleation of the solid discharge product~\cite{Stamm2017,Yin2017}. (2) After the formation of nucleation sites, the precipitation of \ce{ZnO} allows the cell to reach a stable working point. The cell voltage recovers and stabilizes near 1 \si{\volt} as the \ce{Zn} metal electrode continues to dissolve.  (3) When the cell passes a discharged capacity of about 615 $\mathrm{mAh} \cdot \mathrm{g}^{-1}_{\ce{Zn}}$ (75\% \ce{Zn} utilization), the cell voltage decays more rapidly as the available surface area of the \ce{Zn} electrode decreases. Finally, the cell reaches the end of discharge when no usable \ce{Zn} metal remains. 

The complete discharge test confirms that the proposed electrolyte is suitable for primary \ce{Zn}-air batteries at low current densities (0.5 \si{\milli\ampere\per\square\centi\meter}). The ZAB maintains a stable voltage between 0.9 \si{\volt} and 1 \si{\volt} for most of its usable capacity. Furthermore, the ZAB achieves over 98\% of its theoretical capacity at the end of discharging.

The main findings of the full cell electrochemical measurements are as follows. First, the proposed electrolyte is suitable for a primary ZAB, which achieves over 98\% of its theoretical capacity at the end of discharging at a 0.5 \si{\milli\ampere\per\square\centi\meter}. Second, the pH of the electrolyte is stable in the predicted range during cycling. Third, the ZAB cell can be operated with a stable discharging voltage around 1 \si{\volt} for dozens of cycles under various conditions. However, some degradation of the electrolyte is likely to occur during cell charging. 

\subsection{\ce{Zn} electrode characterization}
\label{sec:precipitation}
\begin{figure}[t]
  \includegraphics[width=\columnwidth]{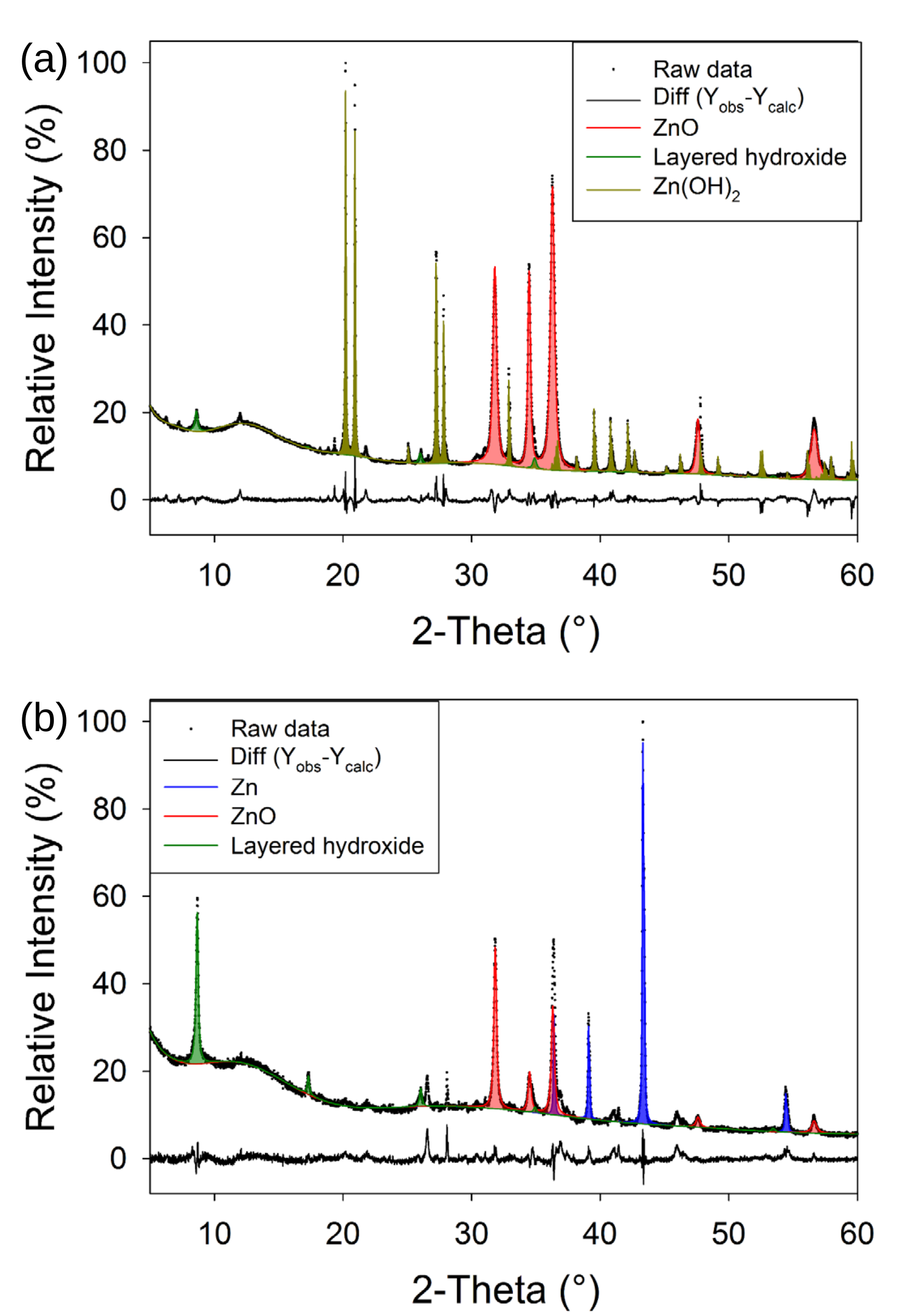}
  \caption{(a) Powder  XRD  pattern  for  anode products formed after 1 full discharge at a rate of  0.5 \si{\milli\ampere\hour\per\square\centi\meter}. (b) Powder  XRD  pattern  for  anode products  formed  after  13  discharge-charge cycles at a rate of 0.5 \si{\milli\ampere\hour\per\square\centi\meter} with a discharging and charging cycle time of 27 hours.  Cycling was stopped on a charging step. Identified phases are fit and displayed in colour.}
  \label{fgr:XRD}
\end{figure}

The \ce{Zn} electrode should form \ce{ZnO} as the final discharge product for the cell to be a true \ce{Zn}-air battery. Existing studies of zinc-based batteries with non-alkaline aqueous electrolytes note that mixed zinc salts - not \ce{ZnO} - dominate the discharge product~\cite{Larcin1997,Lee2016,Clark2019TowardsElectrolytes}. To confirm the presence of \ce{ZnO} in the proposed aqueous organic electrolyte, \ce{Zn} electrodes at various states of charge are characterized using XRD, SEM, and EDS measurements.

Fig. \ref{fgr:XRD} shows the x-ray diffraction patterns collected for two \ce{Zn} electrodes operated under different conditions. After complete discharge (Fig. \ref{fgr:XRD}(a)) two major phases were identified: \ce{ZnO} and \ce{Zn(OH)2}. A small amount of a suspected layered hydroxide phase (discussed in supplementary information$^\dag$) was also observed. Fig. \ref{fgr:XRD}(b) shows the powder XRD pattern for products formed after cycling the cell at 0.5 \si{\milli\ampere\hour\per\square\centi\meter} for 27 \si{\hour} of discharging and 27 \si{\hour} of charging, ending on a charging step. In this charged state, three dominant phases are identified: \ce{Zn}, \ce{ZnO}, and a suspected layered hydroxide phase. The presence of metallic \ce{Zn} can be explained by the mechanism of oxide growth inducing spalling. A shift from \ce{Zn(OH)2} to the layered hydroxide phase is also apparent. It is noted that data fitting was complicated by the observation of considerable preferred orientation in the diffraction pattern of\ce{Zn(OH)2}, and an unknown structural model for the layered hydroxide phase. For this latter material, a simplified model taking only the \{001\} peaks is used. A number of peaks remain unassigned for both analyses. Some of these probably correspond to the layered hydroxide, whilst some are likely due to unidentified minority phases.   

\begin{figure}[t]
  \includegraphics[width=\columnwidth]{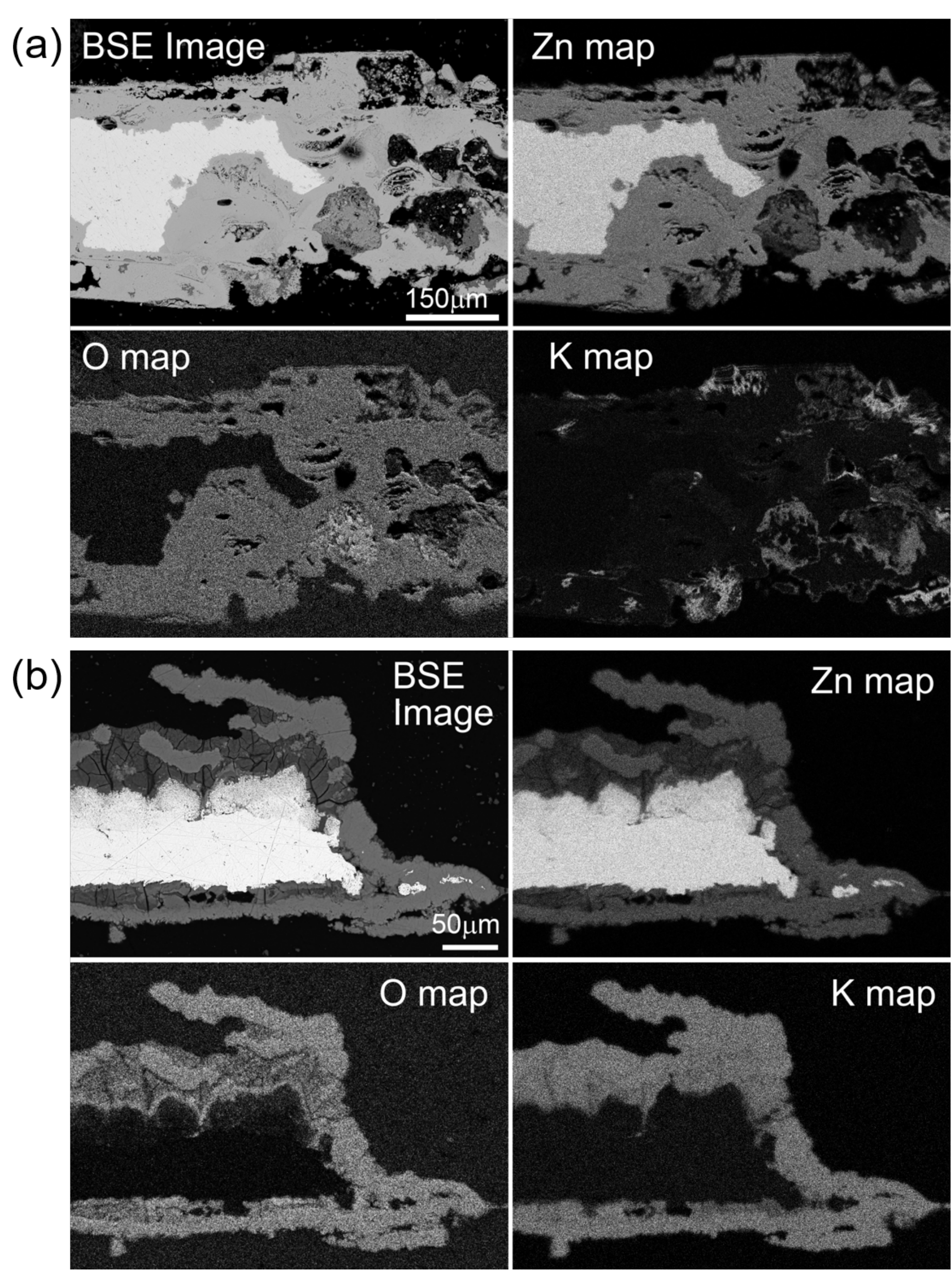}
  \caption{Backscattered electron (BSE) image and EDS analysis of a \ce{Zn} electrode after (a) 1 full discharge at a rate of  0.5 \si{\milli\ampere\hour\per\square\centi\meter} and (b) 13 discharge-charge cycles at a rate of 0.5 \si{\milli\ampere\hour\per\square\centi\meter} with a discharging and charging cycle time of 27 hours. Cycling was stopped on a charging step.}
  \label{fgr:SEM}
\end{figure}

In Fig. \ref{fgr:SEM}, SEM analysis of the electrode cross-sections presents the \ce{Zn} electrode in greater detail. The cross-section images show a complex microstructure. Fig. \ref{fgr:SEM}(a) presents the fully-discharged electrode and Fig. \ref{fgr:SEM}(b) shows the cycled electrode. After discharging, the passivated \ce{Zn} metal that remains is surrounded by a fairly uniform phase rich in \ce{Zn} and \ce{O}. On the other hand, Fig. \ref{fgr:SEM}(b) shows that there are distinct phases present in the cycled \ce{Zn} electrode. The metallic \ce{Zn} phase exhibits a dense layer on the bottom with a more porous \ce{Zn} metal phase on top. The dense layer is the original \ce{Zn} foil, while the porous metal phase is the \ce{Zn} metal deposited during charging. The solids surrounding the \ce{Zn} metal also exhibit clearly distinguishable phases. The layer directly on top of the \ce{Zn} metal contains less \ce{Zn} and \ce{O} than the richer phases towards the outer interface with the electrolyte. This indicates that the solid products are converted back to \ce{Zn} metal first at the electrode interface. However, it is interesting to note that a significant amount of precipitated discharge products remain on the electrode after charging. Additionally, the \ce{Zn} electrode is completely dissolved in some areas. This indicates that the dissolution of the precipitated discharge products during charging may be too slow to support the re-deposition of \ce{Zn} metal. This observation is likely linked to both the slow dissolution kinetics of \ce{ZnO} in neutral solutions~\cite{Zhang1996} and the pH-buffer resisting the shift to more acidic pH values~\cite{Clark2019TowardsElectrolytes}. 

The analysis of \ce{Zn} electrodes in the proposed electrolyte yields two important findings: (i) \ce{ZnO} is a major phase in each sample. The presence of \ce{ZnO} as the dominant discharge product is a significant improvement over existing non-alkaline electrolytes that favor the precipitation of mixed zinc salts and consume electrolyte~\cite{Clark2019TowardsElectrolytes}. The precipitation of \ce{ZnO} allows the battery to achieve a stable working point and opens a pathway towards achieving high energy density. However, (ii) some discharge products remain in the electrode after charging. This indicates that the discharge product dissolution process may be too slow. Additional measurements, discussion, and common \ce{Zn} electrode performance benchmarks~\cite{Stock2019BenchmarkingBatteries} are available in the supplementary information (Figs. S7 \& S8)$^\dag$. 

\section{Conclusions}

The combined theoretical-experimental method of aqueous electrolyte design laid out in this work is applied to develop and test a novel electrolyte composition for rechargeable zinc-air batteries. 

This work proposes thermodynamic descriptors to accelerate the screening of alternative electrolyte materials for zinc-based batteries. The screening shows that there are a wide variety of organic molecules with favorable properties for aqueous ZAB electrolytes. Carboxylic acids and aminocarboxylic acids are particularly suitable. Through the application of theory-based models, we show that a halide-free electrolyte containing citrate and glycine maintains a stable pH during ZAB operation and thermodynamically favors the precipitation of \ce{ZnO} as the final discharge product. Experimental results confirm the validity of the theory-based predictions and yield additional insight into the proposed aqueous-organic ZAB. 

Half-cell RRDE measurements of the air electrode catalyst materials (CNT, EMD, and EMD+CNT) in both nitrogen-saturated and oxygen-saturated electrolyte confirm that these materials are active towards the oxygen reduction reaction, with EMD+CNT showing the greatest activity. However, the measured ring currents indicate that significant quantities of incomplete reaction product \ce{H2O2} are produced via the 2-electron ORR pathway. In the high-potential domain, the current is driven by a combination of organic species oxidation, oxygen evolution, and carbon corrosion.

Full-cell cycling measurements show that the lab-scale cell can be cycled for up to 500 hours. After replacing the electrolyte, the cell is able to recuperate some of its performance and continue cycling. Increasing the volume of the electrolyte and charging at a constant voltage of 2 \si{V} were both shown to give modest improvements in cell cycling lifetime. Finally, the ex-situ SEM, XRD, and EDS measurements confirm that \ce{ZnO} is the dominant discharge product. Analysis of charged \ce{Zn} electrodes show that significant quantities of \ce{ZnO} remain on the electrode surface, indicating that the kinetics of \ce{ZnO} dissolution are slow and could limit the charging time of the battery.

The combined results of the simulation and experiment confirm that the proposed halide-free aqueous electrolyte improves upon state-of-the-art by precipitating \ce{ZnO} and \ce{Zn(OH)2} as the dominant discharge products and maintaining a stable pH during cell operation. However, challenges including the oxidation of the organic molecules during cell charging and slow dissolution of the discharge products remain. Therefore, the proposed electrolyte should be viewed as a proof-of-concept for the design method, and as one formulation of many possibilities. The initial results from experiment validate both the modeling methods applied in the analysis and the underlying understanding that governs them. Continuing to apply these methods to further investigate and optimize halide-free aqueous electrolytes supports the development of novel materials for zinc-based batteries.

\section{Experimental and computational methods}

Three customized electrochemical cell designs with varying distances between the anode and the cathode were used in this work. In the first cell (C1), the electrodes were separated by 0.9 \si{\centi\meter} and 1.1 \si{\milli\liter} of electrolyte was injected. C1 was the default cell, used for all measurements unless otherwise stated. In the second cell (C2), the distance between the electrodes was increased to 2.8 \si{\centi\meter} and 4.4 \si{\milli\liter} of electrolyte was injected. This was done to place two pH microelectrodes (Mettler Toledo, InLab\textsuperscript{\textregistered} Micro) near the positive and negative electrodes for operando pH measurements. In the final cell design (C3), the distance between the electrodes was 1.4 \si{\centi\meter} and 1.85 \si{\milli\liter} of electrolyte was injected. This cell was used to investigate the effect of electrolyte volume on the cycling lifetime of the ZAB. A zinc foil (Alfa Aesar, 99.98\%, 250 \si{\micro\meter} thickness) was used as the Zn electrode in each cell, with an active area of 1.327 \si{\square\centi\meter}. The electrochemical analyses were carried out in a BaSyTEC Battery Test System.

The electrolyte was prepared from dissolving citric acid (Merck Millipore, anhydrous $>$99.5\%) and glycine (Scharlau, reagent grade) in deionized water. The pH value was adjusted to pH=9 with \ce{KOH} (Sigma Aldrich, $>$98\%) and the solution was saturated with \ce{ZnO} (Sigma Aldrich, puriss. p.a.). The as-prepared electrolyte solution contained $[\ce{Cit^{3-}}]_{\mathrm{T}} = 1.80$M, $[\ce{Gly^{-}}]_{\mathrm{T}} = 0.91$M, and $[\ce{Zn^{2+}}]_{\mathrm{T}} = 0.44$M. Physicochemical properties of the as prepared electrolyte system were analyzed with specific equipment for those measurements. In this context, the pH, ionic conductivity (IC), viscosity ($\mu$), dissolved oxygen ([\ce{O2}]), mass density ($\rho$) and total organic carbon (TOC) were measured.

Three catalyst materials were evaluated in this study: carbon nanotubes (CNT, Arkema GraphistrengthTM C100), electrolytic manganese dioxide (EMD, Tosoh Hellas A. I. C.), and a mixture of the two (EMD+CNT). In the three different bifunctional air electrodes 10 wt.\% of PTFE (Dyneon TF 5032 PTFE) was added to the mixture. The CNT bifunctional air electrode and EMD bifunctional air electrode were prepared with 90 wt.\% of CNT and EMD, respectively. On the other hand, the EMD+CNT bifunctional air electrode was composed of 20 wt.\% EMD, 70 wt.\% CNT and 10 wt.\% PTFE. Finally, the mixtures were pressed twice for 1 minute at 50 bar against a carbon gas diffusion layer (Freudenberg H23C9). Once the electrodes were pressed, they were heated at 340 \textsuperscript{$\circ$}C for 30 minutes where 2.2 \si{\milli\gram\per\square\centi\meter} of catalyst loading were achieved. 

The electrochemical measurements were performed with a rotating ring disk electrode (RRDE) setup (Pine Instruments. AFRD-5). A glassy-carbon working electrode (0.196 \si{\square\centi\meter} ) was surrounded by a Pt ring biased at 1.2 \si{\volt}, which allows for simultaneous measurement of the hydrogen peroxide formation in the ORR. A thin film of the catalyst was fabricated onto a mirror polished and cleaned glassy carbon substrate by pipeting an aqueous suspension, drying in the \ce{N2} stream, and pipeting/drying an aqueous suspension of Nafion for fixing the thin film of the catalyst. A Pt wire was used as a counter electrode and a saturated calomel electrode as a reference electrode (all the potentials are referred to a reversible hydrogen electrode, RHE). All chemicals and gasses were of the highest purity from the available choices.

Assessment of the crystalline components of as-tested anodes was performed using a Bruker D8 Advance A25 powder diffractometer equipped with a Cu K-$\alpha$ radiation source and LynxEye XE\textsuperscript{TM} detector (XRD). All measurements were performed in Bragg-Brentano geometry. Data were initially collected on formed \ce{Zn} electrodes, but due to sample inhomogeneity and high levels of anisotropy in the crystallite growth data collection was repeated using reaction product powder scraped from the Zn anode foil. This was grounded and then mounted on a zero background holder (an oriented Si crystal) using silicone grease so as to reduce crystal orientation effects. Phase identification was performed via reference to the ICCD PDF4+ (2017) crystal structure database~\cite{ICDDDatabase} and Crystallographic Open Database (COD)~\cite{Grazulis2009}, and confirmed via whole powder pattern fitting using the Bruker Topas v5 analysis software.

Electron microscopy images were obtained for top surface and cross-sections using an Hitachi S3400N Electron Microscope (SEM) equipped with an Oxford Instruments Aztec EDS system. For cross section analysis, samples were embedded in epoxy resin (Struers EpoFix) and polished down to a fine finish using SiC polishing papers down to 5$\mu$m media size. Sample preparation was performed without water or lubricant in order to avoid sample dissolution. In order to avoid sample charging, top section imaging was performed in Low Vacuum mode, at a chamber pressure of 10 Pa, whilst cross section samples were coated with a thin layer of carbon.

Two types of numerical models are applied in this work. The first is a thermodynamic model of ion speciation and solubility, based on the law of mass action. The modelling method is derived in existing works~\cite{Limpo1993,Limpo1995,Clark2017,Clark2019TowardsElectrolytes}. The second is a cell-level continuum model based on the quasi-particle method derived in our previous works \cite{Clark2017,Clark2019TowardsElectrolytes}. Assuming that the homogeneous acid-base and \ce{Zn}-complexing reactions in the electrolyte occur instantly, the quasi-particle continuum model supports the efficient calculation of the dynamic concentration profiles in the electrolyte and gives insight into the pH stability in the cell during operation. All models with their requisite parameters are described in the supplementary information$^\dag$~\cite{Mainar2018a,Akerlof1941, Davis1967,Sipos2000,Stumpp2015InterplayOxide,Geng2010PredictionPressures,Hawthorne2002Simonkolleite4,Kolodziejczak-Radzimska2014ZincReview,Vazquez-Arenas2012,Latz2013,Amend,Ortiz-Aparicio2007,Marangoni1989a,Hamelers2010,Liu2017a,Amyes2004,Sarada2000,Zirino1972,Moezzi2013FormationDissolution-precipitation,Sobel2008,Apelblat2015,Hamborg2008,Huajun2007,Horstmann2013,Neidhardt2012,Pourbaix1974AtlasSolutions,Clark2018,Clever1992,Iruin2019,Eberle2014,Schmitt2019,Pan2016,Kundu2016,Kundu2018,Wan2018,Garche2009,Hatada1998,Freire2010,Mesmer1973FluoroborateSolutions1a,Wang2018a,Amendola2012,Ortiz-Aparicio2013,Ghose1964TheZn5OH6CO32,Staehlin1970TheZn5OH8NO32.2H2O}. 


\section*{Conflicts of Interest}

The authors S.C., A.M., E.I., L.C., A.B., and B.H. have filed a patent application involving electrolytes for zinc-air batteries.

\section*{Acknowledgement}

This work has received funding from the European Union's Horizon 2020 research and innovation program under grant agreement No. 646186 (ZAS! project), from the Basque Country Government (ELKARTEK 2017 program), and from the Deutsche Forschungsgemeinschaft under project Be 1201/22-1. The support of the bwHPC initiative through the use of the JUSTUS HPC facility at Ulm University is acknowledged. This work contributes to the research performed at CELEST (Center for Electrochemical Energy Storage Ulm-Karlsruhe). The full cell measurements in this work have been done in the frame of the Doctoral Degree Program in Chemistry by the Universitat Aut\`onoma de Barcelona.




\bibliographystyle{advancedmaterials}
\bibliography{references} 

\end{document}